\documentclass[reprint, amsmath, amssymb,aps,prb,superscriptaddress,
floatfix
]{revtex4-2}
\usepackage{graphicx} 
\usepackage{bm}
\usepackage{braket}
\usepackage{mathtools}
\usepackage{color}
\usepackage{amssymb}
\usepackage{amsthm}
\usepackage[dvipsnames]{xcolor}
\definecolor{DarkGreen}{rgb}{0.0, 0.5, 0.0}
\usepackage[colorlinks,citecolor=DarkGreen,linkcolor=blue,linktocpage]{hyperref}
\usepackage{ulem}
\usepackage{comment}
\newcommand{\hk}[1]{{\color[rgb]{1.0,0.1,0.1}#1}}
\newcommand{\so}[1]{{\color{blue}{#1}}}
\newcommand{\soo}[1]{{\color{red}{#1}}}
\newcommand{\sooo}[1]{{\color{red}{#1}}}
\newcommand{\soooo}[1]{{\color{red}{#1}}}
\newcommand{\sooooo}[1]{{\color{red}{#1}}}
\newcommand{\imi}{{\rm i}}
\renewcommand{\hk}{}
\renewcommand{\so}{}
\newcommand{\hatt}{}
\def \kb {}
\def \epsilonnovar {\epsilon}
\renewcommand{\soo}{}
\renewcommand{\sooo}{}
\renewcommand{\soooo}{}
\renewcommand{\sooooo}{}
\def \susy {{^\text{SUSY}}}

\begin{document}
\title{Disorder-free Sachdev-Ye-Kitaev models: 
Integrability and a precursor of chaos}
\author{Soshun Ozaki}
 \affiliation{Department of Basic Science, The University of Tokyo, Komaba, Meguro-ku, Tokyo 153-0041, Japan}
\author{Hosho Katsura}%
\affiliation{Department of Physics, 
The University of Tokyo, 
Hongo, Bunkyo-ku, Tokyo 113-0033, Japan}
\affiliation{Institute for Physics of Intelligence, The University of Tokyo, 
Hongo, Bunkyo-ku, Tokyo 113-0033, Japan}
\affiliation{Trans-scale Quantum Science Institute, The University of Tokyo, 
Hongo, Bunkyo-ku, Tokyo 113-0033, Japan}
\date{\today}

\begin{abstract}
    We introduce two disorder-free variants of the Sachdev-Ye-Kitaev (SYK) model, demonstrate their integrability, and study their static and dynamical properties.
    Unlike diagrammatic techniques, the integrability of these models allows us to obtain  
    dynamical correlation functions even when the number of Majorana fermions is finite.  
    From the solutions, we find that out-of-time-order correlators (OTOCs) in these models exhibit exponential growth at early times, resembling that of many-body quantum chaotic systems such as those with disorder or external kick terms,
    \sooooo{despite their large $N$ behavior differing from that of typical chaotic systems}.  
    Conversely, our analysis shows no evidence of random-matrix behavior in level spacing statistics or the spectral form factor. 
    Our findings illustrate that the clean versions of the SYK models represent
    simple but nontrivial examples of disorder-free quantum many-body systems displaying chaos-like behavior of OTOCs.
\end{abstract}
\maketitle

\section{Introduction}
The nonequilibrium dynamics of quantum many-body systems is one of the central issues in modern physics, encompassing concepts such as thermalization \cite{gogolin2016equilibration, deutsch2018eigenstate, mori2018thermalization}, many-body quantum chaos \cite{d2016quantum, borgonovi2016quantum, liu2020quantum}, quantum scars \cite{serbyn2021quantum, regnault2022quantum, chandran2023quantum}, and many-body localization \cite{nandkishore2015many, alet2018many, abanin2019colloquium}. Recently, there has been renewed interest in many-body quantum chaos, not only within statistical physics but also in high-energy physics and quantum information theory. This resurgence is motivated by the conjecture that black holes act as fast scramblers \cite{hayden,sekino,Shenker2014, Shenker2015, Maldacena2016jhep} and the proposal of the Sachdev-Ye-Kitaev (SYK) model \cite{sy1993, kitaev, trunin, rosenhaus2019introduction, chowdhury2022sachdev}. To date, various metrics have been proposed for diagnosing quantum chaos \cite{kudler2020}. 

Commonly used criteria in analyzing quantum chaotic systems include level spacing statistics and the spectral form factor (SFF). These criteria are based on the conjecture that such systems exhibit properties akin to those of random matrices. Specifically, the level spacing statistics of quantum chaotic systems are expected to follow the Wigner-Dyson distribution, as exemplified in Refs.~\cite{macdonald1979, berry1981}. Additionally, the SFF, interpreted as the two-point correlation function of energy levels, is believed to capture the features of quantum chaos \cite{lau2019,dyer2017}.

In recent years, the out-of-time-order correlators (OTOCs) \cite{larkin1969,Polchinski2016,Maldacena2016prd,Dowling2023} have emerged as another significant indicator of quantum chaos. OTOCs are defined using the quantum counterpart of Poisson brackets in classical mechanics.
\soooo{As in some one-body chaotic systems with periodic kicking, OTOCs in quantum many-body chaotic systems are believed to exhibit scrambling, i.e., exponential growth at early times followed by rapid decay to zero. This behavior is commonly observed in quantum many-body systems with disorder or periodic driving.}
In particular, the exponential growth of OTOCs at early times corresponds to the sensitivity to initial conditions inherent in classical chaotic systems. This exponential growth is quantified by the quantum Lyapunov exponent $\lambda$ \cite{sekino, Maldacena2016jhep}, which has been conjectured to be bounded as $\lambda \leq 2\pi T$ where $T$ is the temperature. This upper bound is referred to as the Maldacena-Shenker-Stanford (MSS) bound. It has been confirmed that some models considered chaotic exhibit this exponential growth in OTOCs \cite{rozenbaum2017kickedroterlyapunov, rozenbaum2019stadium}. 
Note, however, that the OTOC is not a definitive indicator of chaos \cite{hashimoto2020,akutagawa2020,xu2020, Dowling2023}, nor is the SFF \cite{lau2019}. 
\soooo{In this paper 
we say that a quantum many-body system is chaotic if it exhibits Wigner-Dyson-type level spacing statistics, random matrix--like SFF, and the scrambling of OTOCs.} 

These indicators are effective when applied to the SYK model; Level spacing statistics and SFF exhibit random matrix-like behavior \cite{youprb,garciagarciaprd,Cotler2017,GarciaGarcia2018}, while the OTOC shows exponential growth \cite{kitaev, Polchinski2016, Maldacena2016prd, kitaev2018}. In this paper, we show that the OTOC can also detect anomalous dynamics occurring in the disorder-free variants of the SYK model \footnote{See \cite{witten2019syk, klebanov2017uncolored, iyoda2018, krishnan2018contrasting,Balasubramanian2021,Claps} for other disorder-free variants of the SYK models.}, which we dub the clean SYK models. 
\soooo{These models are integrable in the sense that their energy eigenvalues and eigenstates can be obtained explicitly.}
After demonstrating the integrability, we investigate the OTOC\soo{, level spacing statistics,} and SFF
using the obtained solutions. 
We find that the OTOC shows an exponential growth at early times despite the integrability of the models, 
\soooo{although the duration 
is much shorter than that of the typical chaotic systems.}
This behavior may be interpreted as a precursor of chaos. 
In contrast, we observe that the level spacing statistics and SFF of the models do not exhibit random matrix-like behavior.
\soooo{Studying the chaos-reminiscent behavior of the OTOC, which is analytically tractable even for finite $N$, may enhance our understanding of the chaotic dynamics of the (original) SYK model.}

\sooooo{
This paper is organized as follows. In Sec.~\ref{sec:models}, we introduce two disorder-free variants of the SYK model and demonstrate their integrability by finding their analytical solutions explicitly. 
In Sec.~\ref{sec:level_spacing}, we numerically study the many-body density of states and level-spacing statistics of these models. 
In Sec.~\ref{sec:thermodynamics}, we investigate the fundamental thermodynamic properties and discuss the residual entropy.
In Sec.~\ref{sec:dynamics}, we examine dynamical properties. We formulate two-point functions and OTOCs and
evaluate them numerically.
Furthermore, we discuss the exponential behaviors of OTOCs.
In Sec.~\ref{sec:sff}, we numerically evaluate SFF and verify the scaling law obtained from analytical approximation.
Finally, we present a summary in Sec.~\ref{sec:summary}.
}
 
\section{models and solutions\label{sec:models}}
We start with a simple variant of the SYK model 
with uniform coupling \cite{Lau2021}
\begin{align}
    H_4={\rm i}^2\sum_{1\leq i_1 < i_2 < i_3 < i_4 \leq N}
    \gamma_{i_1} \gamma_{i_2} \gamma_{i_3} \gamma_{i_4},
    \label{eq:Ham_H4}
\end{align}
where $\gamma_i$ ($i=1,...,N$) are Majorana fermions satisfying $\gamma^\dagger_i = \gamma_i$ and $\{\gamma_{i_1}, \gamma_{i_2}\}=2\delta_{i_1 i_2}$.
\hk{In order to solve $H_4$, 
we introduce 
the auxiliary quadratic Hamiltonian}
\begin{align}
    H_2 = {\rm i} \sum_{1 \leq i_1 < i_2 \leq N} \gamma_{i_1} \gamma_{i_2}. 
    \label{eq:Ham_H2}
\end{align}
\hk{A straightforward calculation shows that this Hamiltonian and $H_4$ are related to each other via} 
\begin{equation}
    H_4=\frac{1}{2}(H_2)^2+E_0,
    \quad E_0 = -\frac{N(N-1)}{4}.\label{eq:H4_H2_identity}
\end{equation} 
\sooooo{The derivation of Eq.~\eqref{eq:H4_H2_identity} is provided in Appendix~\ref{app:clean_syk_diagonalization}.}
\so{Consequently, diagonalization of $H_4$ is reduced to that of $H_2$.}

To diagonalize $H_2$, we define the creation and annihilation operators for fermions
by the Fourier transform of the Majorana operators: 
\begin{align}
    &f^\pm_k=\frac{1}{\sqrt{2N}} \sum_{j=1}^N e^{\mp {\rm i}(j-1)\theta_k}\gamma_j,
\end{align}
where $\theta_k=(2k-1)\pi/N$ ($k=1,\dots , \frac{N}{2}$), and $f^\pm_k$ satisfy the anticommutation relations: 
\soo{$\{f^{s}_k,f^{\bar{s'}}_{k'}\}=\delta_{ss'}\delta_{kk'}$ with $\bar{s'} (s'=+,-)$ being a short-hand notation for $-s'$.}
Then, we can rewrite $H_2$ 
as
\begin{equation}
    H_2=\sum_{k=1}^\frac{N}{2} \epsilonnovar_k \left( \hatt{n}_k -\frac{1}{2} \right), \quad \epsilonnovar_k=2 \cot\frac{\theta_k}{2},
    \label{eq:h2}
\end{equation}
where $\hatt{n}_k=f^+_k f^-_k$ are the number operators for fermions, each of which takes the values $0$ or $1$.
In this fermionic representation, the number of $k$'s is $N/2$, which is consistent with the fact that the number of Majorana fermions is $N$,
since the degrees of freedom of Majorana fermions are half of those of complex fermions.
Equation~\eqref{eq:h2} indicates that $H_2$ is diagonal in the fermion number basis,
and therefore $H_4$ is.
The energy eigenvalues
of $H_2$ and $H_4$ specified by the occupation numbers $n_1$,
$\dots$, $n_\frac{N}{2}$ are
\begin{equation}
    E_2(n_1, 
    \dots, n_\frac{N}{2})=\frac{1}{2}\sum_{k=1}^\frac{N}{2} (-1)^{n_k+1} 
    \epsilonnovar_k,
    \label{eq:e2}
\end{equation}
and
\begin{equation}
E_4(n_1,
\dots, n_\frac{N}{2})
    =\frac{1}{2} E_2(n_1, 
    \dots, n_\frac{N}{2})^2+E_0,
    \label{eq:e4}
\end{equation}
respectively.  

Unlike in $H_2$, the quasi-particle picture breaks down in $H_4$.
This can be seen by noting that
\begin{align}
    [H_4, f_k^\pm] = (\pm\epsilonnovar_k H_2 - \frac{1}{2} \epsilonnovar_k^2)f_k^\pm,
    \label{eq:diffeq}
\end{align}
where the right-hand side is not linear in $f_k^\pm$
and involves the operator $H_2$. 
\so{This indicates that the energy that the total system gains or loses 
when adding a particle with wave number $k$
depends on the
occupancy of other fermions.
In this sense, a simple quasi-particle picture is not valid for $H_4$.}

We \hk{next}
introduce a clean SYK model with $\mathcal{N}=1$ supersymmetry (SUSY). \hk{The Hamiltonian is}
\begin{equation}
    H_4^{\rm SUSY}=Q^2, \quad Q=\frac{\rm i}{\sqrt{N}} \sum_{1\leq i_1 < i_2< i_3\leq N} \gamma_{i_1} \gamma_{i_2}\gamma_{i_3},
    \label{eq:h4susydef}
\end{equation}
where $Q$ is the supercharge that satisfies $Q^\dagger=Q$ and anticommutes with the fermionic parity: $\{Q,(-1)^F\}=0$.
\so{This model is a disorder-free variant of the SUSY SYK model \cite{sachdevprd2017susy}.}
\so{The model has a Majorana zero mode}
\begin{align}
    &\chi_0=\frac{1}{\sqrt{N}}\sum_{j=1}^N \gamma_j,
    \label{eq:chi0}
\end{align}
which satisfies $(\chi_0)^2=1$ and $[Q, \chi_0]=0$.
To diagonalize $H_4^{\rm SUSY}=Q^2$, it is convenient to factorize $Q$ as
\begin{equation}
    Q=\chi_0 H_{\rm free}, \quad H_{\rm free}= \frac{\rm i}{2}\sum_{j,k} 
    \gamma_j \tilde{\mathcal{A}}_{jk} \gamma_k,
    \label{eq:Ham_Hfree}
\end{equation}
where $\tilde{\mathcal{A}}$ is a real skew-symmetric matrix \hk{whose elements are}
$\tilde{\mathcal{A}}_{jk}=1-\frac{2|k-j|}{N}$ for $j<k$.
\sooooo{The details of this calculation can be found in Appendix~\ref{app:clean_SUSY_SYK_diagonalization}.}
Since $\chi_0$ and $H_{\rm free}$ commute, $H_4^{\rm SUSY}$ is 
written as the square of $H_{\rm free}$, i.e.,
\begin{equation}
    H_4^{\rm SUSY}=(H_{\rm free})^2.
    \label{eq:H4SUSY_H2free_identity}
\end{equation}
Therefore, $H_4^{\rm SUSY}$ is diagonal 
in the basis where
$H_{\rm free}$ is diagonal.
Similarly to the case of $H_4$, $H_{\rm free}$ is rewritten as
\begin{equation}
    H_{\rm free}= \sum_{k=1}^{\frac{N}{2}-1}
    \varepsilon_k \left(\hatt{m}_k -\frac{1}{2} \right), \quad \varepsilon_k=2\cot \frac{\vartheta_k}{2},
\end{equation}
where $\hatt{m}_k=g_k^+ g_k^-$, $\vartheta_k=2k\pi/N$, and $g_k^\pm$ are the fermionic creation/annihilation operators defined by
the 
Fourier transform
\begin{align}
    &g_k^\pm=\frac{1}{\sqrt{2N}} \sum_{j=1}^N e^{\mp {\rm i}(j-1)\vartheta_k} \gamma_j
\end{align}
for $1\leq k\leq N/2-1$.
They satisfy \soo{$\{g^s_k,g^{\bar{s'}}_{k'}\}=\delta_{ss'}\delta_{kk'}$}.
Many-body energy eigenvalues \hk{of $H_4^{\rm SUSY}$} specified by the occupation numbers $m_1$, 
$\dots$, $m_{\frac{N}{2}-1}$ are
\begin{equation}
    E_4^{\rm SUSY}(m_1, \dots, m_{\frac{N}{2}-1})=\frac{1}{4} \left(\sum_{k=1}^{\frac{N}{2}-1}
    (-1)^{m_k+1}\varepsilon_k\right)^2.
    \label{eq:e4susy}
\end{equation}
This expression involves only $N/2-1$ fermions. 
This is because the Majorana zero modes $\chi_0$ and 
\begin{equation}
    \chi_{N/2}=\frac{1}{\sqrt{N}}\sum_{j=1}^N (-1)^{j-1} \gamma_j
    \label{eq:zeromode2}
\end{equation}
do not appear in the Hamiltonian.
We can define two distinct vacua
$\ket{0}_+$ and $\ket{0}_-$ such that $g^-_k\ket{0}_\pm=0$ for each $k$ and $(-1)^F\ket{0}_\pm=\pm\ket{0}_\pm$. 
They are related by $\chi_0$ since $\{ \chi_0, (-1)^F\} = [H_4^{\rm SUSY}, (-1)^F]=0$. Note that the degeneracy of these states is protected by supersymmetry. 
In this way, the missing degrees of freedom are restored.
We can construct a tower of states by applying the creation operators $g^+_k$
to each vacuum.

\hk{We remark that} $H_4$ and $H^{\rm SUSY}_4$ can be mapped \hk{to} classical Ising models 
\so{with all-to-all antiferromagnetic interactions.}
The spin Hamiltonian corresponding to $H_4$ is
\begin{equation}
    H_4=\sum_{k_1,k_2=1}^{N/2} J_{k_1,k_2}\sigma_{k_1}\sigma_{k_2}+
    E_0,
\end{equation}
where $J_{k_1,k_2}=\frac{1}{2} \cot (\theta_{k_1}/2) \cot (\theta_{k_2}/2)$ and 
$\sigma_{k}=2\hatt{n}_k-1$. 
From the form of the Hamiltonian, it is obvious that each $\sigma_{k}$ commutes with $H_4$. 
A similar representation is obtained for $H_4^{\rm SUSY}$.
The integrability originates from the structure that the Hamiltonian consists of the product of conserved quantities.
This structure is commonly found in several other integrable models \cite{jaynes-cummings,hatsugai-kohmoto, xian}.

\section{Level-spacing statistics}
\label{sec:level_spacing}
\sooooo{Having computed the spectra of $H_4$ and $H_4^{\rm SUSY}$, we now study their level-spacing statistics. Let $E_i$ 
($i=1,2,...,2^\frac{N}{2}$)
denote the many-body energy eigenvalues in ascending order.
The many-body density of states (DOS) is defined by
\begin{equation}
    D(E)=\frac{1}{2^\frac{N}{2}} \sum_{i=1}^{2^\frac{N}{2}} \delta(E-E_i). 
\end{equation}}

Next, we examine the level-spacing statistics.
In general, the mean level spacing in a system varies with energy. To enable meaningful comparisons across different energy ranges and 
systems, we adjust the energy spectrum locally---stretching or compressing it---so that the density of states becomes uniform within the energy range of interest \cite{haake_unfolding,tezuka_unfolding}.
We also remove the degeneracy since it can affect the mean level spacing.
After these adjustments,  
we compute the level-spacing distribution $P(s)$, which is defined as the probability density for two consecutive energy levels, normalized by the mean level spacing $\Delta$, to have a spacing $s$.

The numerical results of the many-body DOS 
and level spacing statistics for $H_4$ and $H_4^{\rm SUSY}$,
obtained using the Lorentzian representation of the delta function $\delta(x)=\delta/\pi(x^2+\delta^2)$ with broadening $\delta=0.1$, are shown in \sooo{Fig.~\ref{fig:doslevelstat}}.
\sooooo{Magnified views of the DOS are shown in Fig.~\ref{fig:dos_h2} in Appendix~\ref{app:many_body_DoS}. For comparison, the DOS for $H_2$ and $H_\text{free}$ are also shown in Appendix~\ref{app:many_body_DoS}.}
\begin{figure}
    \centering
    \includegraphics[width=\linewidth]{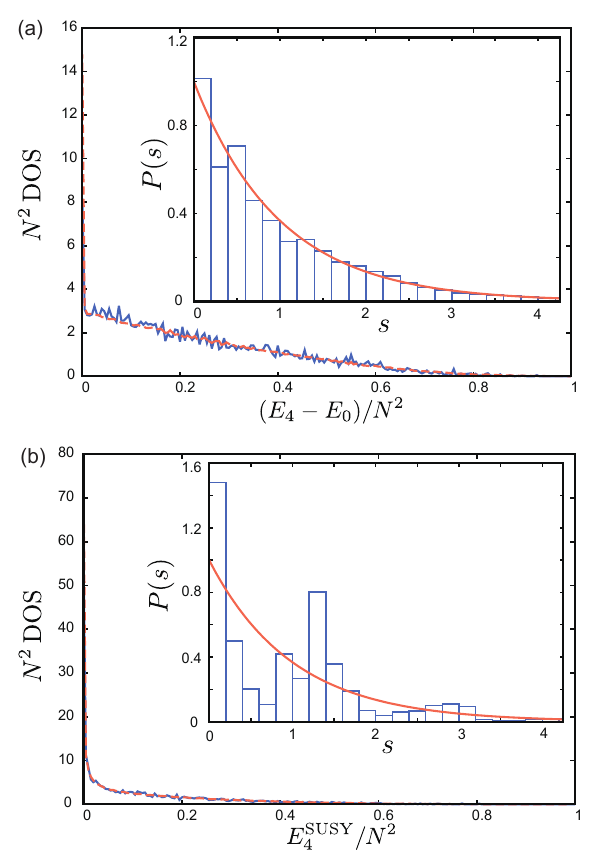}
    \caption{\sooo{Density of states for (a) $H_4$ and (b) $H_4^{\rm SUSY}$ with $N=32$ (solid blue line) and $N=44$ (dashed orange line).
    Each inset 
    shows the corresponding level spacing statistics for $N=44$, with $s$ being the normalized level spacing.  
    The energy eigenvalues in the energy windows $0.4N^2\leq E_4 \leq 0.5 N^2$ and
    $0.2N^2 \leq E_4^{\rm SUSY} \leq 0.3N^2$ are used for 
    $H_4$ and $H_4^{\rm SUSY}$, respectively. 
    The curve $P(s) = e^{-s}$ (orange) is shown for comparison. 
    We have reproduced the data for (a) for completeness, although similar results were obtained in Ref.~\cite{Lau2021}.}}
    \label{fig:doslevelstat}
\end{figure}

For $H_4$, we find that the many-body DOS has a sharp peak at $E=E_0$ and that the level-spacing distribution follows a Poisson law \cite{guhr1998}, which is typically observed in integrable systems. 
This result is consistent with Ref.~\cite{Lau2021}.
The many-body DOS of $H_4^{\rm SUSY}$ displays a divergent behavior towards $E=0$ and also shows level concentration, 
which is in contrast to the sharp peak at $E=E_0$ observed for $H_4$. 
The level-spacing distribution for $H_4^{\rm SUSY}$ is found to be Poisson-like. 

\sooooo{We now analyze the many-body DOS for $H_4$ near the ground state in more detail. Let $W(\Delta E)$ be the number of energy levels within the range $[E_0,E_0+\Delta E]$. The results of $W(\Delta E)$ for different values of $N$ and $\Delta E$ are summarized in Table~\ref{table:statesnum}. We find that these values can be approximated by
\begin{equation}
    W(\Delta E)\simeq a\times 2^{\frac{N}{2}}\sqrt{\soo{\Delta E}},
\end{equation}
with $a\simeq 0.007$.
This result indicates that the approximate ground-state degeneracy grows exponentially with $N$. In the next section, we argue that the presence of a large number of states near the ground state is responsible for the residual entropy in the zero-temperature limit.}
\begin{table*}[htb]
    \caption{Number $W(\Delta E)$ of energy levels in the range $[E_0,
    E_0+\Delta E]$ for $H_4$. \label{table:statesnum}}
    \begin{ruledtabular}
    \begin{tabular}{crrrrrr}
     &  $W(10^{-4})$ & $W(10^{-3})$ & $W(10^{-2})$ & $W(0.02)$ & $W(0.05)$ & $W(0.1)$  \\ \hline
    $N=36$ & 22  & 58 & 186 & 262 & 420 & 592\\
    $N=40$ & 74  & 236 & 730 & 1038 & 1640 & 2342\\
    $N=44$ & 282 & 922 & 2822 & 4032 & 6474 & 9032\\
    \end{tabular}
    \end{ruledtabular}
\end{table*}

\section{Thermodynamic properties\label{sec:thermodynamics}}
\so{
In this section, we discuss the thermodynamic properties of the models. 
We denote the inverse temperature 
by $\beta=1/T$.}
We first compute the partition function of $H_4$, which is defined by $Z(\beta)={\rm Tr} e^{-\beta H_4}$. A more explicit expression of $Z(\beta)$ that is convenient for numerical evaluation at large $N$ can be obtained by the Hubbard-Stratonovich (HS) transformation as
\begin{equation}
    Z\soo{(\beta)}=z\soo{(\beta)} e^{-\beta E_0}\sqrt{\frac{2}{\pi\beta}}2^{\frac{N}{2}}, \label{eq:partfunc}
\end{equation}
where
\begin{equation}
    z\soo{(\beta)}=\int_{-\infty}^\infty e^{-\frac{2x^2}{\beta}} p_N(x) dx,\quad p_N(x)=\prod_{k=1}^{N/2} \cos (\epsilon_k x).
    \label{eq:smallz}
\end{equation}
\sooooo{See Appendix~\ref{sec:sff-fintemp} and 
for the details of the derivation.}

\begin{figure}
    \centering
    \includegraphics[width=\linewidth]{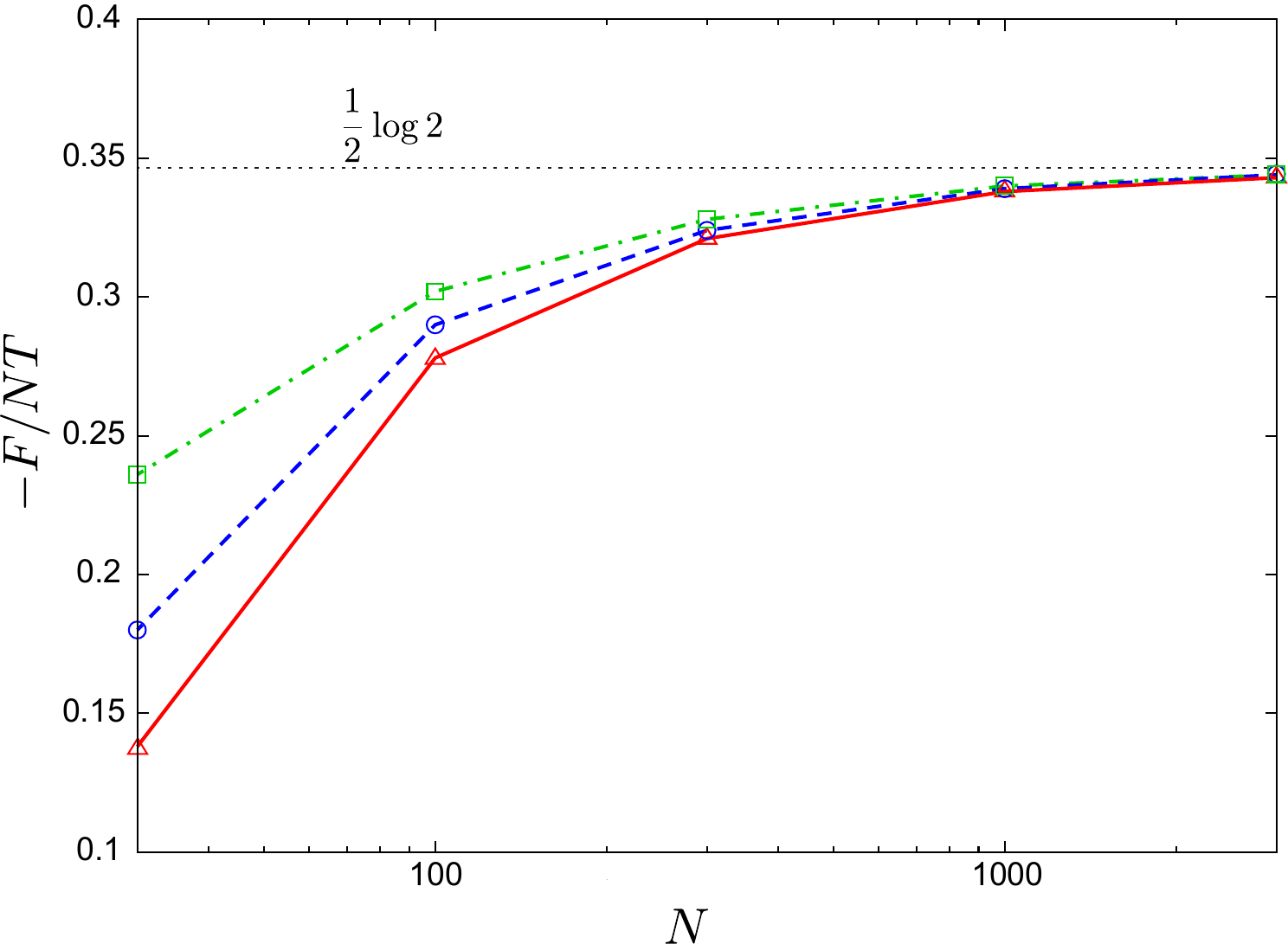}
    \caption{Free energy of $H_4$ as a function of the total number of Majorana fermions \hk{$N$}.
    The solid (red), dashed (blue), and dot-dashed (green) lines 
    correspond to temperature $T=0.1$, $1.0$, and $10.0$.
    The constant shift by $E_0$ is omitted.}
    \label{fig:freeene}
\end{figure}

The above integral can be evaluated numerically.
Figure~\ref{fig:freeene} shows 
$-F\soo{(\beta)}/(NT)=\log Z\soo{(\beta)}/N$, where $F\soo{(\beta)}$ is the free energy.

At each temperature, 
the free energy approaches $\frac{N}{2}\kb T \log 2$
as $N$ increases,
which suggests that the system has nonzero residual entropy.
This is also understood from the fact that $z\soo{(\beta)}$ in Eq.~\eqref{eq:partfunc} is of the order of $O(\frac{1}{N})$ 
since $p_N(x)$ in $z\soo{(\beta)}$ [Eq.~\eqref{eq:smallz}] is approximated as
\begin{equation}
p_N(x)\sim \exp\left(\frac{N}{2}\int_0^\infty \frac{\log|\cos x\epsilon|}{1+(\epsilon/2)^2} 
\frac{d\epsilon}{\pi} \right)
\sim e^{-N|x|} \label{eq:pnapprox}
\end{equation}
in the large $N$ limit.
These results lead to a residual entropy of $s=\frac{1}{2}\log 2$, 
reflecting the presence of a large number of energy levels near the ground state, 
as discussed in the previous section.
The finite residual entropy is also found in the original SYK model,
which is slightly lower ($s_{\rm SYK}\sim 0.2324{\kb}=\frac{1}{2}\log(1.592){\kb} $\cite{Maldacena2016jhep,Cotler2017}) than 
in the present model.
This discussion also applies to $H_4^{\rm SUSY}$, leading to the same residual entropy as $H_4$. See Appendix \ref{sec:app_thermodynamics} for details. 

\section{Two-point functions and OTOCs \label{sec:dynamics}}
\begin{figure}
    \centering
    \includegraphics[width=\linewidth]{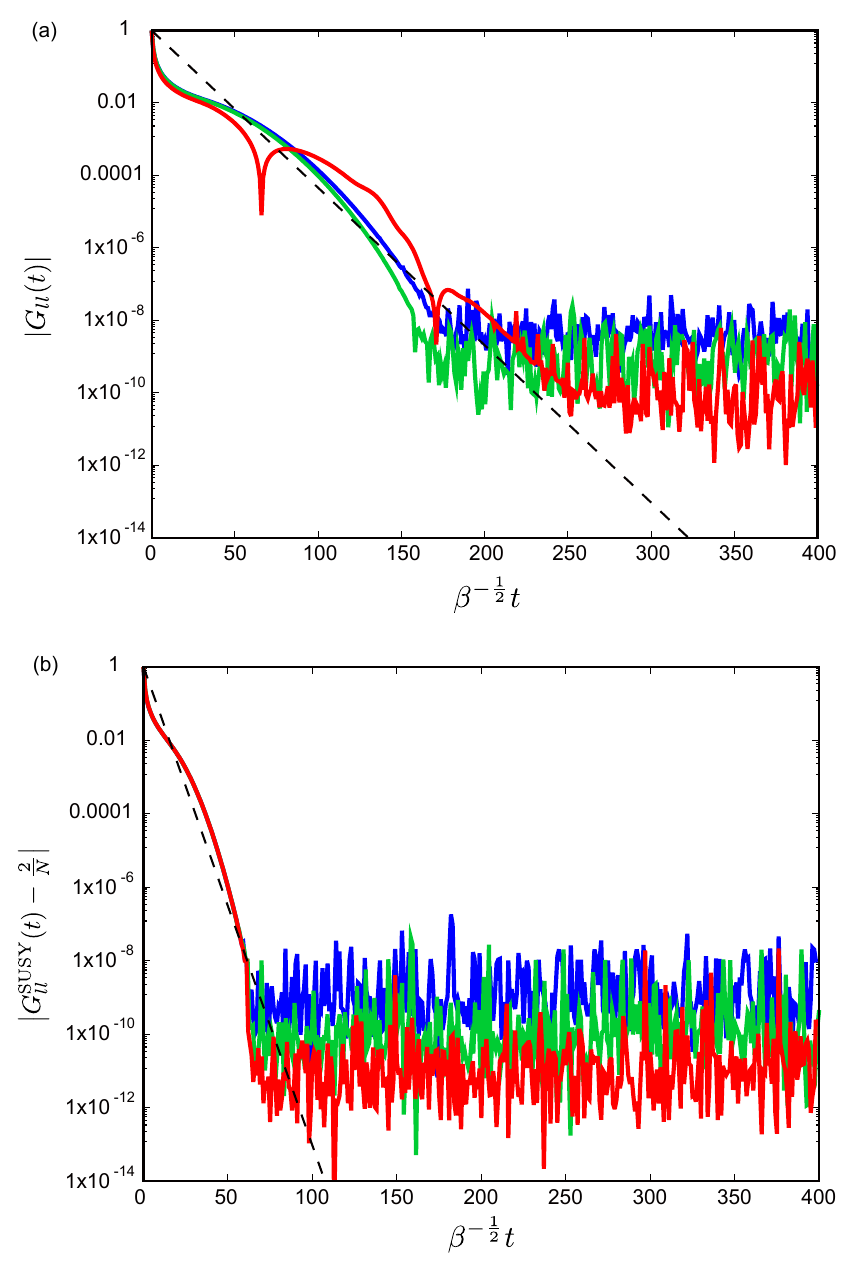}
    \caption{Autocorrelation functions (a) $G_{ll}(t)$ and (b) $G_{ll}^{\rm SUSY}(t)$ at $N=100$. The red, green, and blue lines represent the cases with $\beta=0.01$, $0.1$, and $1$.
    The dashed lines in (a) and (b) represent the scaling $e^{-0.1t/\sqrt{\beta}}$ and $e^{-0.3t/\sqrt{\beta}}$, respectively.}
    \label{fig:autocorr}
\end{figure}

In this section, we study the dynamical properties of $H_4$.
Using Eq.~\eqref{eq:diffeq}, the equation 
of motion of $f_k^\pm(t):=e^{{\rm i} H_4 t} f_k^\pm e^{-{\rm i} H_4 t}$ 
can be integrated analytically \cite{ozaki-nakazato}.
As a result, we obtain an explicit expression for the time evolution of $\gamma_j$: 
\begin{align}
    \gamma_j (t) :&= e^{\imi H_4 t} \gamma_j e^{-\imi H_4 t} \nonumber \\
    &=\sqrt{\frac{2}{N}} \sum_{s=\pm} \sum_{k=1}^{N/2}
    e^{ \imi s(j-1)\theta_k} e^{-\frac{\imi}{2}\epsilon_k^2t} e^{\imi s\epsilon_k H_2 t} 
    f_k^s. \label{eq:gammat}
\end{align}
Similar relations 
can be derived for $H_4^{\rm SUSY}$, which are shown in Appendix~\ref{sec:app_dynamics}. 
We first consider the two-point function
\begin{equation}
    G_{lm}(t)= {\rm Tr}[\rho\, \gamma_l(t) \gamma_m(0)],
    \label{eq:autocorr}
\end{equation}
where $\rho=\exp(-\beta H_4)/{\rm Tr}\exp(-\beta H_4)$ is the density matrix in thermal equilibrium at inverse temperature $\beta$.
Substituting Eq.~\eqref{eq:gammat} to Eq.~\eqref{eq:autocorr}, we obtain
\begin{align}
    G_{lm}(t) 
    =\frac{2}{N} {\rm Tr} \left[\rho \sum_{k=1}^{N/2} 
    e^{-\frac{\imi}{2}\epsilonnovar_k^2 t} e^{{\rm i}(\theta_k(l-m) +\epsilonnovar_k H_2 t)\sigma_k} \right].
\end{align}
The two-point function for $H_4^\text{SUSY}$, $G_{lm}^\text{SUSY}(t)$, is defined similarly.
\sooooo{Figures~\ref{fig:autocorr} (a) and (b) show the numerical results for the autocorrelation functions $G_{ll}(t)$ and $G^\text{SUSY}_{ll}(t)$, corresponding to the case $l=m$.}
The precise definition of $G^\text{SUSY}_{lm}(t)$ and the numerical method we use is presented in Appendix~\ref{app:autoco}.

Numerical observation suggests that $G_{ll}(t)$ and $G_{ll}^{\rm SUSY}(t)$ decay exponentially as $e^{-t/t_{\rm d}}$, up to a constant, as seen in the original SYK model~\cite{Polchinski2016}, and then saturate after a long time.
For $H_4^{\rm SUSY}$, we find that $G_{ll}^{\rm SUSY}(t)$ does not completely decay and has a residual value $2/N$ due to the Majorana zero modes $\chi_0$ [Eq.~\eqref{eq:chi0}] and $\chi_{N/2}$ [Eq.~\eqref{eq:zeromode2}] which do not evolve in time. See Appendix~\ref{app:autoco} for details.
This saturation time tends to become longer as $N$ increases.
The auto-correlation times are obtained as $t_{\rm d}\simeq \sqrt{\beta}/0.1$ for $H_4$ and $t_{\rm d}\simeq \sqrt{\beta}/0.3$ for $H_4^\text{SUSY}$, which are 
different from $t_{\rm d}\sim \beta$
found in the original SYK model \cite{Polchinski2016}.

We now turn to consider the OTOC \cite{Maldacena2016prd,Maldacena2016jhep,Polchinski2016, GarciaGarcia2018}
\begin{equation}
    F(t,\beta)= \frac{1}{N^2}\sum_{l,m=1}^N{\rm Tr} \left[ \rho^\frac{1}{4} \gamma_l(t)\rho^\frac{1}{4} \gamma_m(0)\rho^\frac{1}{4} \gamma_l(t) \rho^\frac{1}{4}\gamma_m(0) \right].
    \label{eq:otoc}
\end{equation}
\so{Similarly to the case of two-point functions, we substitute Eq.~\eqref{eq:gammat} to Eq.~\eqref{eq:otoc} and 
carry out the summation over $l$ and $m$. 
After some algebra [See Appendix~\ref{app:otoc} for details], we obtain}
\begin{align}
    F(t,\beta)=&-\frac{4}{N^2} \sum_{k_1\neq k_2} e^{-\frac{\beta}{4}(\epsilonnovar_{k_1}^2+\epsilonnovar_{k_2}^2)} \nonumber \\
    &\times{\rm Tr} [\rho e^{\frac{\beta}{2}(\sigma_{k_1}\epsilonnovar_{k_1}+\sigma_{k_2}\epsilonnovar_{k_2})H_2}
    e^{({\rm i}t-\frac{\beta}{4})\sigma_{k_1}\sigma_{k_2}\epsilonnovar_{k_1}\epsilonnovar_{k_2}}].
    \label{eq:otocgeneric}
\end{align}
We can calculate the OTOC for $H_4^{\rm SUSY}$ similarly, and find that its large $N$ limit 
takes the form of Eq.~\eqref{eq:otocgeneric} up to a constant of the order of $O(\frac{1}{N})$, with the substitutions $t\to 2t$ and $\beta\to 2\beta$ \soo{[See Appendix~\ref{app:otoc} for details].}
\hk{Below we only show} the results of OTOC for $H_4$.

First, we consider the infinite temperature limit ($\beta=0)$,
where the density matrix is proportional to the identity matrix.
To compute $F(t, \beta)$ in Eq.~\eqref{eq:otocgeneric}, we use 
$\frac{2}{N}\sum_{k=1}^{N/2}(\cdots)=\int_0^\infty d\epsilonnovar\rho(\epsilonnovar)(\cdots)$ with $\rho(\epsilonnovar)=\frac{1}{\pi}[1+(\frac{\epsilonnovar}{2})^2]^{-1}$.
\so{As a result, we obtain}
\begin{align}
    F(t,0)&=-\frac{2}{\pi}f(4t) + O\left( \frac{1}{N} \right), \nonumber\\
    &\sim 
    \begin{dcases}
        -1+\frac{8}{\pi}(1-\gamma-\log 4t)t + O(t^2) & (t\ll 1)\\
        -\frac{1}{2\pi t} + O\biggl(t^{-2}\soo{,\frac{1}{N}}\biggr) & (t \gg 1),
    \end{dcases}
\end{align}
where $f(x)=\int_0^\infty \sin(u)/(u+x)du$ is one of the auxiliary function 
for the trigonometric integrals
and $\gamma=0.5772\cdots$ is the Euler-Mascheroni constant.
We find that the time derivative of the OTOC at $t=0$ diverges and that the OTOC 
grows 
rapidly \so{at early times}.
This 
rapid growth is reminiscent of those in (weakly) chaotic systems \cite{kukuljan2017,prakash2020}.
Note that the MSS bound $\lambda \leq 2\pi T$ at infinite temperature does not prohibit the divergent growth of OTOC.
At late times, on the other hand, the OTOC shows $t^{-1}$ power law decay.
Such a power-law decay of OTOC is commonly seen in some integrable 
models \cite{lin2018,bao2020} and the bipartite kicked rotor model
\cite{prakash2020}.

Next, we consider 
the OTOC at finite temperature. 
By the HS transformation, we can calculate the trace in Eq.~\eqref{eq:otocgeneric} and obtain
\begin{align}
 F(t,\beta)=&-\frac{4}{zN^2}\sum_{k_1\neq k_2} e^{-\frac{\beta}{8}
    (\epsilonnovar_{k_1}^2+\epsilonnovar_{k_2}^2)}
    \cos (\epsilonnovar_{k_1} \epsilonnovar_{k_2}t) \nonumber \\
    &\times \int_{-\infty}^\infty dx  e^{-\frac{2x^2}{\beta}}
    \prod_{k'\in K\backslash\{k_1,k_2\}}\cos x \epsilonnovar_{k'}.
\end{align}
\begin{figure}
    \centering
    \includegraphics[width=\linewidth]{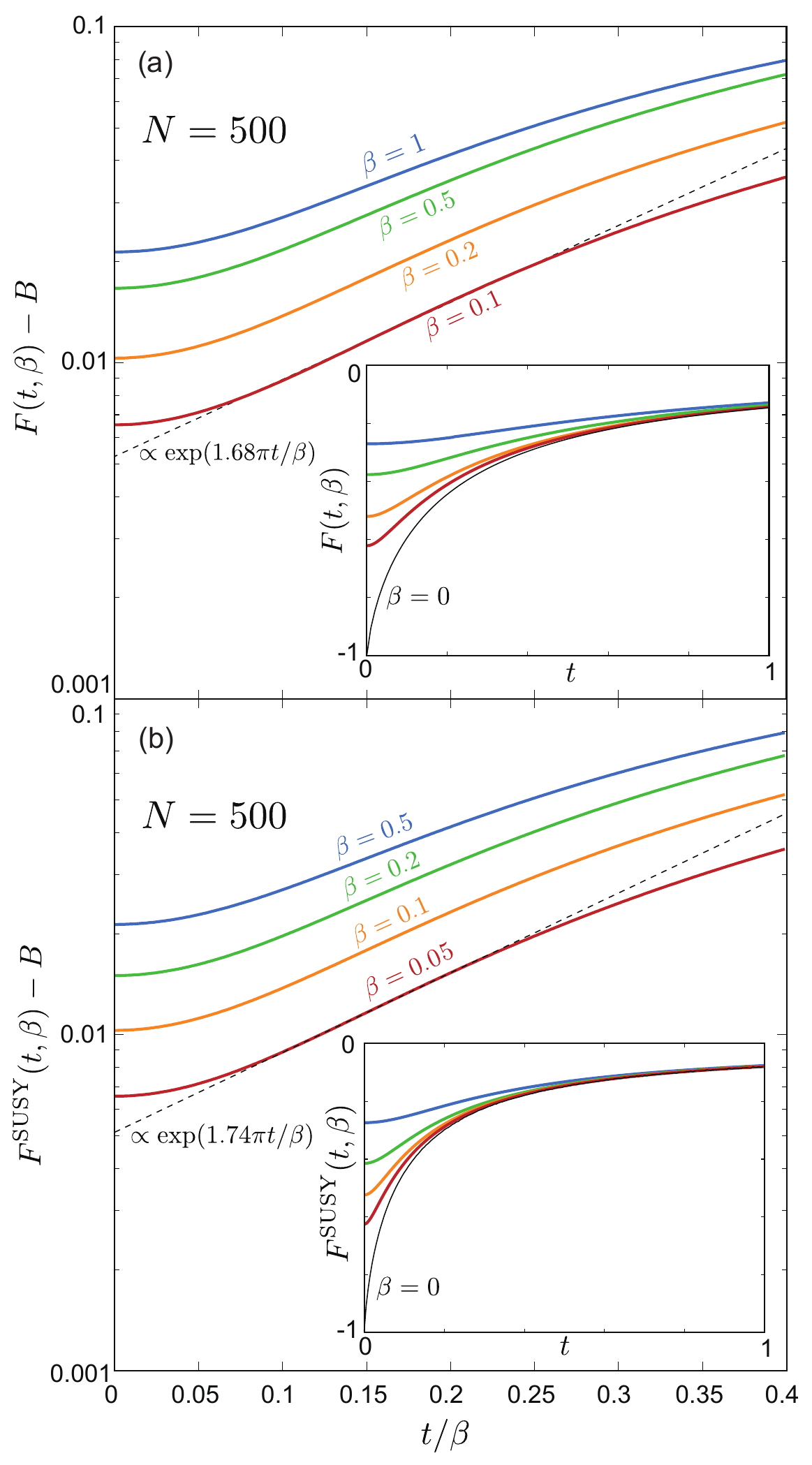}
    \caption{
    \soo{OTOCs for (a) $H_4$ and (b) $H_4^{\rm SUSY}$ for various temperatures as functions of \soooo{the rescaled time $t/\beta$}, each with a constant $B$ 
    subtracted. The constant $B$ is obtained by fitting the function 
    $F_{\rm fit}(t)=A e^{\lambda t}+B$ to numerical data
    over [$0.1 \beta$,$0.2\beta$].
    Note that $A$ and $B$ depend on $\beta$.
    Insets: OTOCs without subtraction.}
    }
    \label{fig:otoc_h4}
\end{figure}
\noindent
\soo{
The insets of Fig.~\ref{fig:otoc_h4} show \sooooo{the OTOCs for $H_4$, $F(t,\beta$), and for $H_4^{\rm SUSY}$, $F^{\rm SUSY}(t,\beta),$} as functions of time for different values of $T=1/\beta$, evaluated by numerical integration. 
We find that the OTOCs exhibit exponential-like growth at early times.
Assuming that 
this behavior is well described by the function
$F_{\rm fit}(t)=Ae^{\lambda t}+B$,
we fit the parameters $A$, $\lambda$, and $B$ to the results.
The fitting is carried out in the interval $[0.1\beta,0.2\beta]$ for each temperature
and the exponential parts ($F(t,\beta)-B=Ae^{\lambda t}$) are shown in Fig.~\ref{fig:otoc_h4}.}
\soooo{
Although a polynomial fit is possible, we find that the exponential fit yields more reasonable fitting parameters. See Appendix~\ref{app:otoc} for the details of the fitting.}
The exponents of the observed exponential-like growth are comparable to the MSS bound $2\pi T$, 
which may be a precursor of quantum chaos.
\soooo{This exponential behavior of the OTOC is remarkable for disorder-free integrable systems without kicking/driving, although saturation of OTOCs toward zero is widely seen, e.g., see Ref.~\cite{Dowling2023}.}

\soooo{
We observe that the duration of the exponential growth of the present model is about 
$\beta/5$ and almost independent of $N$, which is quite different from $\beta \log N$, the typical scrambling time of chaotic systems \cite{kobrin_syk_numerical}.
This may imply that a simple operator in the present model only grows to a small size in the operator space.}
As expected, this exponential-like behavior of the OTOC 
is absent in $H_2$ or $H_{\rm free}$,
where the OTOC is $-1$ for any time $t$ and does not show any scrambling behavior in the large $N$ limit [See Appendix~\ref{app:otoc}]. 

\section{Spectral form factor \label{sec:sff}}
Finally, in this section, we examine the SFF.
For $H_4$, the SFF is defined by
\begin{equation}
    g(t,\beta)=\left| \frac{{\rm \,Tr} e^{({\rm i}t-\beta) H_4}}{{\rm \,Tr} e^{-\beta H_4}} \right|^2,
    \label{eq:def_SFF}
\end{equation}
where we do not 
perform disorder averages since the models we consider are free of disorder. 
For simplicity, we consider the case of $\beta=0$.
\sooooo{The early-time behavior of $g(t,0)$ can be computed analytically, as discussed in Appendix \ref{sec:sff-smallt}.} We are, however, more interested in the late-time asymptotic behavior.
Using the HS transformation, we obtain
\begin{align}
    g(t,0)
    =&\frac{2}{t\pi} \left|
    \int_{-\infty}^\infty 
    e^{-{\rm i}\frac{2x^2}{t}} p_N(x) dx \right|^2.\label{eq:sff}
\end{align}
For large $N$,
as Eq.~\eqref{eq:pnapprox} suggests, $p_N(x)$ defined in Eq.~\eqref{eq:smallz} has non-negligible values for $|x| \lesssim 1/N$.
This fact is verified in Fig.~\ref{fig:pfuncs} in Appendix~\ref{sec:sff-largeN}.
In this region, the oscillating factor can be regarded as unity for $N^2t\gg1$.
A similar expression is obtained for $H_4^{\rm SUSY}$.
Furthermore,
we find numerical evidence that the scaling
$p_{k N}(x)\simeq p_N(k x)$ holds for a positive integer $k$ and large $N$.
From this relation, we can extract the scaling of SFF,
$g(t,0)\sim C/N^2 t$, where the constant $C$ is determined by 
numerical integration: $0.16$ for $H_4$ and $1.9$ for $H_4^{\rm SUSY}$.
[See Appendix~\ref{sec:sff-largeN} for details.]

Figure~\ref{fig:sff-many} (a) and (b) show the SFF for $H_4$ and $H_4^{\rm SUSY}$, respectively, numerically obtained from Eq.~\eqref{eq:sff} for $H_4$ and from its counterpart for $H_4^{\rm SUSY}$ [Eq.~\eqref{eq:h4susy-sff-int}] for finite $N$.
\soooo{The inset shows the SFF for $H_4$ 
over a longer period of time for $N=48$.}
\so{
We find that the SFF is self-averaged and that the numerical results 
approach the obtained scaling law at large $t$ as $N$ increases.}
\soooo{Shortly before the inverse mean level spacing $\tau\sim 2^{N/2}/N^2$ (approximately $7\times 10^3$ at $N=48$), the SFF begin oscillating.}
These results do not 
have a random matrix interpretation. 

\begin{figure}
    \centering
    \includegraphics[width=\linewidth]{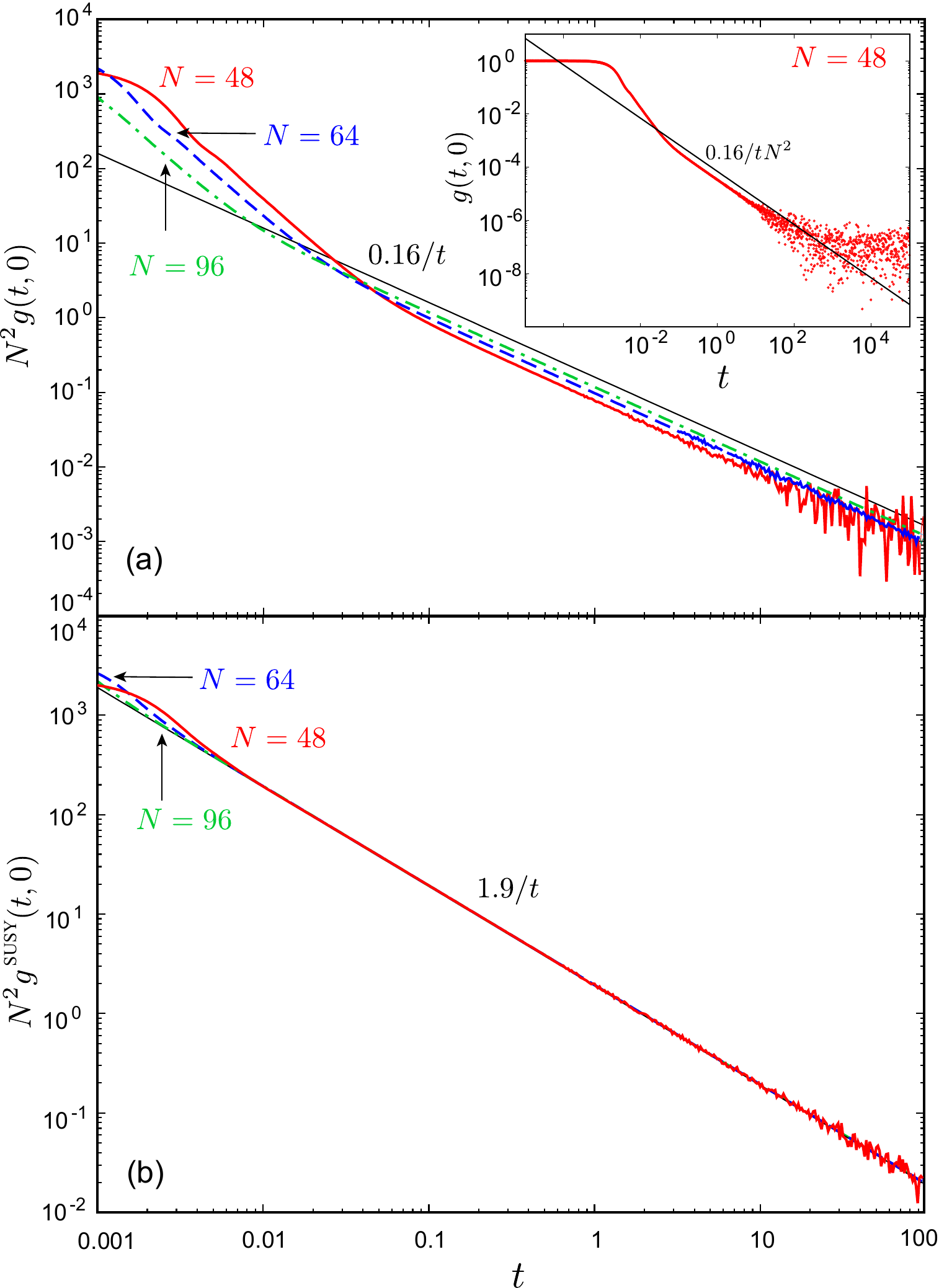}
    \caption{SFF 
    for (a) $H_4$ and (b) $H_4^{\rm SUSY}$ as functions of $t$. The solid (red), dashed (blue), and dot-dashed (green) lines represent the 
    results for $N=48$, $64$, and $96$, respectively. They approach $0.16/t$ with increasing $N$.
    \soooo{Inset: The SFF 
    for $H_4$ over a longer period of time for $N=48$.}}
    \label{fig:sff-many}
\end{figure}

\section{summary \label{sec:summary}}
We have introduced two variants of clean SYK models and demonstrated their integrability. By exploiting this integrability, we have investigated 
the static and dynamical properties of the models. Notably, these models exhibit exponential behaviors in OTOCs at early times, akin to those observed in certain quantum chaotic systems,
\soooo{although the duration 
is much shorter than that for typical chaotic systems.}
\soo{This behavior can be interpreted as a precursor of chaos.} 
Unlike typical chaotic systems that involve disorder or external kick terms, our models do not incorporate such elements.
As anticipated, on the other hand, our analysis revealed no evidence of 
random matrix behavior in the level spacing statistics or SFF, 
indicating the integrable nature of these models. Consequently, our findings illustrate that clean SYK models 
provide notable examples of disorder-free quantum many-body systems displaying chaos-like behavior of OTOCs.

\begin{acknowledgments}

We are grateful for fruitful discussions with Neil Dowling, Eiki Iyoda,  Atsushi Iwaki, Hiroki Nakai, Pratik Nandy, Ryotaro Suzuki, and Masaki Tezuka.
H.K. was supported by JSPS KAKENHI Grants No.\ JP18K03445, No.\ JP23K25790, No.\ JP23K25783, and MEXT KAKENHI Grant-in-Aid for Transformative Research Areas A “Extreme Universe” (KAKENHI Grant No.\ JP21H05191).
S.O. was supported by JSPS KAKENHI Grant No.\ JP22KJ0988. 

\end{acknowledgments}

\appendix

\section{Many-body eigenenergies and eigenstates of the Clean SYK models
\label{sec:eigenval}}
In this appendix, we show the detailed derivation of the energy eigenvalues and eigenstates of the clean SYK Hamiltonian $H_4$ in Eq.~\eqref{eq:Ham_H4} and the clean SUSY SYK Hamiltonian $H_{\rm SUSY}$ in Eq.~\eqref{eq:h4susydef}.

\subsection{Clean SYK model}
\label{app:clean_syk_diagonalization}
Here we prove Eq.~\eqref{eq:H4_H2_identity}.
Consider the square of $H_2$:
\begin{equation}
    (H_2)^2=\imi ^2 \sum_{i<j} \sum_{k<l} \gamma_i \gamma_j \gamma_k \gamma_l.
\end{equation}
The double sum can be simplified by considering 
the ordering of the indices $i$, $j$, $k$, and $l$. Each term is summarized in Table~\ref{table:fourmajorana}.
\begin{table}
    \caption{Product of Majorana operators with different order  
    of the indices $i$, $j$, $k$, and $l$.}
    \begin{ruledtabular}
    \begin{tabular}{ccc}
   \#  &  ordering of the indices & $\gamma_i \gamma_j \gamma_k \gamma_l$\\ \hline
    1 & $i<j<k<l$  & $\gamma_i \gamma_j \gamma_k \gamma_l$\\
    2 & $i<j=k<l$  & $\gamma_i\gamma_l$\\
    3 & $i<k<j<l$  & $-\gamma_i \gamma_k \gamma_j \gamma_l$\\
    4 & $i<k<j=l$  & $-\gamma_i\gamma_k$\\
    5 & $i<k<l<j$  & $\gamma_i\gamma_k \gamma_l \gamma_j$\\
    6 & $i=k<j<l$  & $-\gamma_j\gamma_l$\\
    7 & $i=k<j=l$  & $-1$\\
    8 & $i=k<l<j$  & $\gamma_l\gamma_j$\\
    9 & $k<i<j<l$  & $\gamma_k\gamma_i \gamma_j \gamma_l$\\
    10 & $k<i<j=l$  & $\gamma_k\gamma_i$\\
    11 & $k<i<l<j$  & $-\gamma_k \gamma_i \gamma_l \gamma_j$\\
    12 & $k<i=l<j$  & $-\gamma_k\gamma_j$\\
    13 & $k<l<i<j$  & $\gamma_k \gamma_l \gamma_i \gamma_j$\\
    \end{tabular}
    \end{ruledtabular}
    \label{table:fourmajorana}
\end{table}
Most of the terms in the table cancel each other, and we are left with $\# 1$, $\# 7$, and $\# 13$. 
As a result, we obtain the desired Eq.~\eqref{eq:H4_H2_identity}. 

It is clear from Eq. \eqref{eq:H4_H2_identity} that the basis that diagonalizes $H_2$ also diagonalizes $H_4$. The quadratic Hamiltonian $H_2$ is diagonalized as follows \cite{Lau2021}.
We rewrite $H_2$ in a matrix form
\begin{equation}
    H_2 = \frac{\imi}{2}\sum_{jk}\mathcal{A}_{jk}\gamma_j \gamma_k,
\end{equation}
where $\mathcal{A}$ is an $N\times N$ real skew symmetric matrix, whose components are
\begin{equation}
    \mathcal{A}_{jk}=
    \begin{cases}
        1 & \text{($j<k$)} \\
        0 & \text{($j=k$)} \\
        -1 & \text{($j>k$)}
    \end{cases}.
\end{equation}
Since the matrix $\mathcal{A}$ is a circulant matrix with 
anti-periodic boundary conditions,
$\mathcal{A}$ is diagonalized by
\begin{equation}
    \bm{u}_k=\frac{1}{\sqrt{N}}
    \begin{pmatrix}
        1, &  e^{\imi\theta_k}, & e^{\imi 2\theta_k}, & \cdots, & e^{\imi(N-1)\theta_k} 
    \end{pmatrix}^{\rm T}
\end{equation}
with $k=1, 2, \cdots, N$ and $\theta_k=(2k-1)\pi/N$.
The eigenvalue of $\mathcal{A}$ corresponding to $\bm{u}_k$ is obtained as
$\imi \cot\frac{\theta_k}{2}$. 
Using $\bm{u}_k$, we can construct an orthogonal matrix $\mathcal{O}$ such that 
\begin{align}
    \mathcal{O}^{\rm T} \mathcal{A} \mathcal{O} 
    = \bigoplus^{\frac{N}{2}}_{k=1} 
    \begin{pmatrix}
        0 & \cot\frac{\theta_k}{2} \\
        -\cot\frac{\theta_k}{2} & 0 
    \end{pmatrix}. 
\end{align}
Explicitly, it is given by
\begin{widetext}
\begin{align}
    \mathcal{O}&=\left( \frac{\bm{u}_1+\bm{u}_N}{\sqrt{2}}, \frac{\bm{u}_1-\bm{u}_N}{\sqrt{2}\imi},
        \frac{\bm{u}_2+\bm{u}_{N-1}}{\sqrt{2}}, \frac{\bm{u}_2-\bm{u}_{N-1}}{\sqrt{2}\imi}, \cdots,
        \frac{\bm{u}_{\frac{N}{2}}+\bm{u}_{\frac{N}{2}+1}}{\sqrt{2}},
        \frac{\bm{u}_{\frac{N}{2}}-\bm{u}_{\frac{N}{2}+1}}{\sqrt{2}\imi}\right) \nonumber \\
    &=\sqrt{\frac{2}{N}} \begin{pmatrix}
        1 & 0 & \cdots & 1 & 0 \\
        \cos\theta_1 & \sin\theta_1 & \cdots & \cos\theta_{\frac{N}{2}} & \sin\theta_{\frac{N}{2}} \\
        \vdots & \vdots & & \vdots & \vdots \\
        \cos (N-1)\theta_1 & \sin (N-1)\theta_1 & \cdots  &
        \cos(N-1)\theta_{\frac{N}{2}} & \sin(N-1)\theta_{\frac{N}{2}} \\
    \end{pmatrix},
\end{align}
\end{widetext}
where we have used the relation $\bm{u}_{N-k+1}=\bm{u}_k^*$.

This transformation defines a new set of Majorana fermions 
\begin{align}
    \psi_{2k-1} &=\sqrt\frac{2}{N}\sum_{j=1}^N \cos[(j-1)\theta_k] \gamma_j, \\
    \psi_{2k} &=\sqrt\frac{2}{N}\sum_{j=1}^N \sin[(j-1)\theta_k] \gamma_j,
\end{align}
in terms of which the quadratic Hamiltonian $H_2$ is expressed as
\begin{equation}
    H_2 = \frac{\imi}{2}\sum_{k=1}^\frac{N}{2} \cot \frac{\theta_k}{2}(\psi_{2k-1} \psi_{2k} - \psi_{2k} \psi_{2k-1}).
\end{equation}
To rewrite $H_2$ in a diagonal form, we introduce 
\begin{equation}
    f_k^\pm =\frac{\psi_{2k-1} \mp \imi \psi_{2k}}{2} 
    \label{eq:complexf}
\end{equation}
with anticommutation relations $\{f_k^{s_1},f_{k'}^{\bar{s}_2}\}=\delta_{s_1,s_2}\delta_{k,k'}$, where $\bar{s}$ ($s=+$, $-$) is a short-hand notation for $-s$. In terms of complex fermions, $H_2$ takes the form Eq.~\eqref{eq:h2}, which leads to the energy eigenvalues
$E_2(n_1, \dots, n_\frac{N}{2})$ [Eq.~\eqref{eq:e2}] and $E_4(n_1, \dots, n_\frac{N}{2})$ [Eq.~\eqref{eq:e4}]
where $n_k=0$ or $1$ is the occupation number for the fermion 
corresponding to $\epsilon_k$.
In this fermionic representation, the maximum number of $k$'s is $k_{\rm max}=\frac{N}{2}$.

\subsection{Clean SUSY SYK model}
\label{app:clean_SUSY_SYK_diagonalization}
The Hamiltonian and the supercharge defined in Eq.~\eqref{eq:h4susydef} satisfy the $\mathcal{N}=1$ supersymmetry algebra
\begin{equation}
    Q^\dagger=Q,\quad\{ Q,(-1)^F\}=0.
\end{equation}
Here $(-1)^F={\rm i}^{N/2} \gamma_1 \gamma_2 \cdots \gamma_N$ is the fermionic parity.  
We find a Majorana zero mode $\chi_0$ [Eq.~\eqref{eq:chi0}]
that commute with the Hamiltonian $H_4^{\rm SUSY}$, which follows from
$[Q, \chi_0]=0$.
Using $\chi_0$, we factorize $Q$ as shown below: 
\begin{widetext}
\begin{align}
    Q=&(\chi_0)^2Q \nonumber \\
    =&\chi_0 \frac{\imi}{N} \sum_{l=1}^N\sum_{1\leq i<j<k \leq N} \gamma_l \gamma_i \gamma_j \gamma_k 
    \nonumber \\
    =&\chi_0 \frac{\imi}{N}  \bigg(
    \sum_{1\leq l<i<j<k\leq N} +\sum_{1\leq i<l<j<k\leq N} +\sum_{1\leq i<j<l<k\leq N} +\sum_{1\leq i<j<k<l\leq N} \bigg) \gamma_l \gamma_i \gamma_j  \gamma_k 
    \nonumber \\
    &+\chi_0 \frac{\imi}{N} \sum_{1\leq i<j<k\leq N} \left[ \gamma_i \gamma_i\gamma_j \gamma_k + \gamma_j \gamma_i\gamma_j \gamma_k+\gamma_k \gamma_i\gamma_j \gamma_k \right].
\end{align}
\end{widetext}
Since the Majorana fermions anticommute on different sites, the first term vanishes. The second term (the term in the last line) is simplified as $ \chi_0 H_{\rm free}$, or equivalently, $H_{\rm free} \chi_0$ with 
\begin{align}
H_{\rm free}= \frac{\imi}{2} \sum_{j,k} 
    \tilde{\mathcal{A}}_{jk} \gamma_j \gamma_k, \quad
\tilde{\mathcal{A}}_{jk}=
    \begin{cases}
        1-\frac{2|k-j|}{N} & \text{($j<k$)} \\
        0 & \text{($j=k$)} \\
        -1+\frac{2|k-j|}{N} & \text{($j>k$)}
    \end{cases}.
    \label{eq:hfreesusy}
\end{align}
Note that the commutation relation $[\chi_0, H_\text{free}]=0$ follows from $[Q,\chi_0]=0$.
To sum up, we have $Q=\chi_0 H_{\rm free}$ and
obtain a simple relation Eq.~\eqref{eq:H4SUSY_H2free_identity} between $H_4^{\rm SUSY}$ and $H_{\rm free}$,
which is analogous to Eq. (\ref{eq:H4_H2_identity}). 
The Hamiltonian $H_4^{\rm SUSY}$ consists of the square of three-body interaction; however, Eq.~\eqref{eq:H4SUSY_H2free_identity} indicates that $H_4^{\rm SUSY}$ contains four-body terms at most. Furthermore, the quadratic terms vanish in $H_4^\text{SUSY}$ due to the antisymmetric property of the matrix $\tilde{\mathcal{A}}$ in Eq. (\ref{eq:hfreesusy}). Therefore, $H_4^{\rm SUSY}$ consists only of the quartic terms up to a constant.

We now diagonalize $H_{\rm free}$. Since $\tilde{\mathcal{A}}$ is a circulant matrix with 
periodic boundary conditions, $\tilde{\mathcal{A}}$ is diagonalized by the following eigenvectors,
\begin{equation}
    \bm{v}_k=\frac{1}{\sqrt{N}}
    \begin{pmatrix}
        1, &  e^{\imi\vartheta_k}, & e^{\imi2\vartheta_k}, & \cdots, & e^{\imi(N-1)\vartheta_k} 
    \end{pmatrix}^{\rm T},
\end{equation}
for $k=0, 1, \dots, N-1$ 
with $\vartheta_k=2k\pi/N$, and the corresponding eigenvalues are obtained as 
\begin{equation}
    \lambda_k=
    \begin{cases}
        0 & (k=0 \,\text{or} \,k=\frac{N}{2}) \\
        {\rm i}\cot \frac{\vartheta_k}{2} & \text{(otherwise)}
    \end{cases}.
\end{equation}
This result implies another Majorana zero mode $\chi_{N/2}$ [Eq.~\eqref{eq:zeromode2}],
corresponding to ${\bm v}_{N/2}$ as well as $\chi_0$, corresponding to ${\bm v}_0$.
Similarly to the case of the clean SYK model, 
the orthogonal matrix $\tilde{\mathcal{O}}$ defined by
\begin{widetext}
\begin{align}
    \tilde{\mathcal{O}}&=\left( \frac{\bm{v}_1+\bm{v}_{N-1}}{\sqrt{2}}, \frac{\bm{v}_1-\bm{v}_{N-1}}{\sqrt{2}\imi},
        \frac{\bm{v}_2+\bm{v}_{N-2}}{\sqrt{2}}, \frac{\bm{v}_2-\bm{v}_{N-2}}{\sqrt{2}\imi}, \cdots,
        \frac{\bm{v}_{\frac{N}{2}-1}+\bm{v}_{\frac{N}{2}+1}}{\sqrt{2}},
        \frac{\bm{v}_{\frac{N}{2}-1}-\bm{v}_{\frac{N}{2}+1}}{\sqrt{2}\imi},
        \bm{v}_0, \bm{v}_{\frac{N}{2}}\right), 
\end{align}
\end{widetext}
reduces $\tilde{\mathcal{A}}$ to a canonical form:
\begin{align}
    \tilde{\mathcal{O}}^{\rm T} \tilde{\mathcal{A}} \tilde{\mathcal{O}}
    = \left( 
        \bigoplus^{\frac{N}{2}-1}_{k=1}
             \begin{pmatrix}
                0 & \cot \frac{\vartheta_k}{2}\\
                -\cot \frac{\vartheta_k}{2} & 0
             \end{pmatrix}
      \right) 
      \oplus \begin{pmatrix}
                0 & 0 \\
                0 & 0
             \end{pmatrix}.
\end{align}
Consequently, we can rewrite the Hamiltonian as
\begin{equation}
    H_{\rm free}^{\rm SUSY}= \sum_{k=1}^{\frac{N}{2}-1}
    \varepsilon_k \left(g_k^+ g_k^- -\frac{1}{2} \right), \quad \varepsilon_k=2\cot \frac{\vartheta_k}{2},
    \label{eq:hfreefermion}
\end{equation}
where 
\begin{align}
    g_k^\pm=\frac{\chi_{2k-1} \mp \imi \chi_{2k}}{2} \quad \text{for}\, k=1,\dots \frac{N}{2}-1\label{eq:complexg}
\end{align}
with
\begin{align}
    \chi_{2k-1}&=\sqrt{\frac{2}{N}} \sum_{j=1}^N \cos[(j-1)\vartheta_k] \gamma_j, \\
    \chi_{2k}&=\sqrt{\frac{2}{N}} \sum_{j=1}^N \sin[(j-1)\vartheta_k] \gamma_j,
\end{align} 
obeying the anticommutation relations 
$\{g_k^{s_1},g_{k'}^{\bar{s}_2}\}=\delta_{s_1,s_2}\delta_{k,k'}$.
It then follows from Eq. (\ref{eq:H4SUSY_H2free_identity}) that the many-body energy eigenvalue of $H_4^{\rm SUSY}$ is given by
$E_4^\text{SUSY}(m_1, \dots, m_{\frac{N}{2}-1})$ [Eq.~\eqref{eq:e4susy}], where  $m_k=0$ or $1$ is the occupation number for the fermion corresponding to $\varepsilon_k$. In this fermionic representation, the maximum number of $k$'s is $k_\text{max}=\frac{N}{2}-1$.

\subsection{Many-body DOS and level statistics}
\label{app:many_body_DoS}
\begin{figure*}
    \centering
    \includegraphics[width=\linewidth]{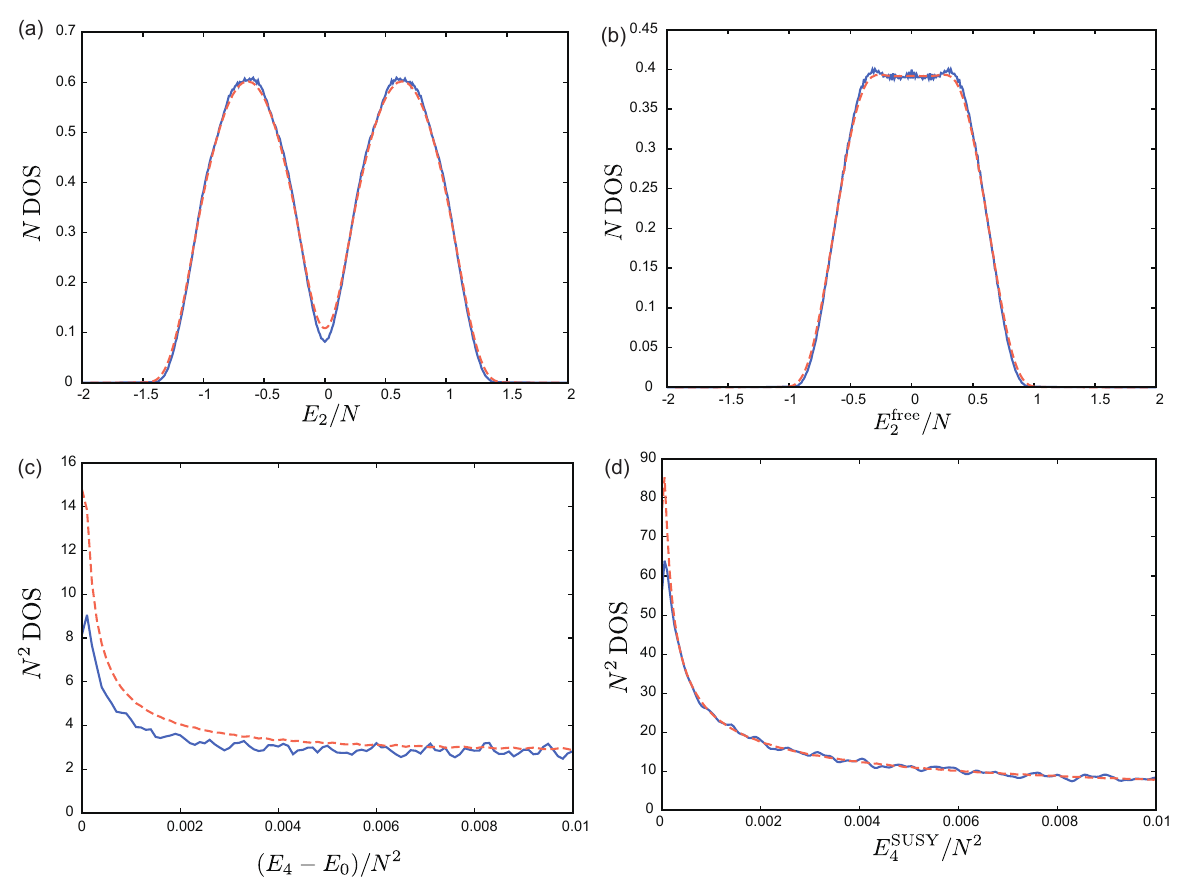}
    \caption{
    Density of states for (a) $H_2$, (b) $H_\text{free}$, (c) $H_4$, and (d) $H_4\susy$
    with $N=32$ (solid blue line) and $N=44$ (dashed orange line).
    $E_2$, $E_2^{\rm free}$, $E_4$, and $E_4^{\rm SUSY}$ represent the energy eigenvalues for $H_2$, $H_{\rm free}$, $H_4$, and $H_4^{\rm SUSY}$, respectively.
    Note that the normalizations are different.
    Panels (c) and (d) present the magnified views of Fig.~\ref{fig:doslevelstat} (a) and (b), respectively.
    \label{fig:dos_h2}}
\end{figure*}

The many-body DOS for $H_2$, $H_\text{free}$, $H_4$, and $H_4^{\rm SUSY}$ are shown in Figs.~\ref{fig:dos_h2}(a), (b), (c), and (d), respectively, with magnified views provided for the latter two.
The many-body DOS for $H_2$ exhibits a double-peak structure, whereas that for $H_{\rm free}$ shows a flat peak.
The squared Hamiltonians $H_4$ and $H_4^{\rm SUSY}$ demonstrate a significant level concentration near the ground state. 

\section{Spectral form factors\label{sec:app_sff}}
In this 
appendix, we show the detailed calculations of the 
SFF of $H_4$ and $H_4^\text{SUSY}$. 
The SFF of $H_4$ is defined in Eq. \eqref{eq:def_SFF}. To proceed, it is more convenient to rewrite it as
\begin{equation}
    g(t,\beta)
    =\left| \frac{{\rm \,Tr} e^{(\imi t-\beta) (H_4-E_0)}}{Z_0(\beta)} \right|^2,
    \label{eq:sff_append}
\end{equation}
where 
\begin{equation}
    Z_0(\beta)={\rm Tr} e^{-\beta (H_4-E_0)} = {\rm Tr} e^{-\frac{\beta}{2}(H_2)^2}. \label{eq:partfunc0}
\end{equation}
The partition function of $H_4$, i.e., ${\rm Tr} e^{-\beta H_4}$, and $Z_0(\beta)$ are related to each other by $Z(\beta)=e^{-\beta E_0} Z_0(\beta)$.
The SFF of $H_4^\text{SUSY}$ is defined as
\begin{equation}
    g^\text{SUSY}(t,\beta)=\left| \frac{{\rm \,Tr} e^{(\imi t-\beta) H_4^\text{SUSY}}}{Z^\text{SUSY}(\beta)} \right|^2,
    \label{eq:sff-susy}
\end{equation}
where 
\begin{equation}
Z^\text{SUSY}(\beta)={\rm Tr} e^{-\beta H_4^\text{SUSY}} = {\rm Tr} e^{-\beta (H_\text{free})^2}.
\label{eq:partfuncsusy}
\end{equation}

 We show the results for early-time and large $N$ behavior at the infinite tempeature in Sec.~\ref{sec:sff-smallt} and \ref{sec:sff-largeN}, respectively. In these cases, we have $Z(\beta=0)=Z^\text{SUSY}(\beta=0)=2^\frac{N}{2}$. 
We also give expressions for SFFs at finite temperature in Sec.~\ref{sec:sff-fintemp}.

\subsection{Small $t$ expansion \label{sec:sff-smallt}}
Here, we show that the power series expansion of $g(t,\beta=0)$ up to $t^2$ is given by
\begin{equation}
    g(t,0)=1-\binom{N}{4} t^2 + O(t^4),
    \label{eq:g_t_short_time}
\end{equation}
which indicates that $g(t,0)$ decays faster at early times with increasing $N$.
To derive the above expression, we expand the numerator in Eq.~\eqref{eq:sff_append} in $t$ as
\begin{align}
    {\rm Tr}\, e^{\imi t H_4}
    =&1+ \imi t {\rm Tr}\, H_4
    +\frac{(\imi t)^2}{2!} {\rm Tr}\, (H_4)^2
    +\frac{(\imi t)^3}{3!} {\rm Tr}\, (H_4)^3 \nonumber \\
    &+O(t^4).
\end{align}
We show below that
\begin{align}
    {\rm Tr}\, H_4 &= 0,\\
    {\rm Tr}\, (H_4)^2 & = 2^{\frac{N}{2}} \binom{N}{4}, 
    \label{eq:Tr_H4^2} \\
    {\rm Tr}\, (H_4)^3 & = 2^{\frac{N}{2}} \cdot 6 \binom{N}{6}.
    \label{eq:Tr_H4^3}
\end{align}

To evaluate these traces, it is convenient to use another expression of the Hamiltonian: 
\begin{equation}
    H_4=\frac{1}{4} \sum_{a< b} \sigma_{k_a} \sigma_{k_b} \epsilon_{k_a} \epsilon_{k_b},
\end{equation}
where $\sigma_{k_a}:= 2n_{k_a}-1$.
This expression leads to ${\rm Tr} H_4=0$.
This also follows from the original expression Eq.~\eqref{eq:Ham_H4} using the cyclicity of the trace and the anti-commutation relation of Majorana fermions. 

Next, we evaluate ${\rm Tr} (H_4)^2$. In terms of the spin variables, it can be expressed as
\begin{widetext}
\begin{equation}
    {\rm Tr} (H_4)^2 = \frac{1}{4^2}\sum_{\sigma_1, \cdots, \sigma_{N/2}=\pm 1} \sum_{a<b} \sum_{c<d}
    \sigma_{k_a} \sigma_{k_b} \sigma_{k_c} \sigma_{k_d} 
    \epsilon_{k_a}\epsilon_{k_b}\epsilon_{k_c}\epsilon_{k_d}.
\end{equation}
In the sum, the terms that satisfy $a=c$ and $b=d$ only remain, which leads to Eq. \eqref{eq:Tr_H4^2}. 

In the third order, similarly, we have
\begin{equation}
    {\rm Tr} (H_4)^3 = \frac{1}{4^3}\sum_{\sigma_1,\dots,\sigma_{N/2}=\pm 1} \sum_{a<b} \sum_{c<d} \sum_{e<f}
    \sigma_{k_a} \sigma_{k_b} \sigma_{k_c} \sigma_{k_d} \sigma_{k_e} \sigma_{k_f}
    \epsilon_{k_a}\epsilon_{k_b}\epsilon_{k_c}\epsilon_{k_d} 
    \epsilon_{k_e} \epsilon_{k_f}.
\end{equation}
\end{widetext}
\clearpage
The terms that satisfy one of the following conditions
\begin{equation}
    \begin{cases}
        a=c<b=e<d=f \\
        a=c<d=e<b=f \\
        a=e<b=c<d=f \\
        a=e<c=f<b=d \\
        c=e<a=d<b=f \\
        c=e<a=f<b=d.
    \end{cases}
\end{equation}
give nonzero contributions and other terms vanish after taking the summation over $\sigma_1, \dots, \sigma_{N/2}$.
Therefore, we obtain 
\begin{align}
    {\rm Tr}(H_4)^3 &= \frac{2^6}{4^3} 6\cdot 2^\frac{N}{2} \sum_{1\leq a<b<c \leq \frac{N}{2}}
    \prod_{i=a,b,c}
    \cot^2 \biggl(\frac{2k_i-1}{2N}\pi\biggr). 
    \label{eq:sff3rd}
\end{align}

This expression can be simplified using the following identity which holds  for $1\leq j \leq n$:
\begin{equation}
    \sum_{1\leq i_1<i_2< \cdots < i_j \leq n} \prod_{m=1}^j \cot^2 \left(\frac{2k_{i_m}-1}{4n}\pi \right)
    = \binom{2n}{2j}
    \label{eq:sumformula}
\end{equation}

\medskip

\noindent
\begin{proof}
We begin with the identity
\begin{equation}
    (e^{\imi \theta_j})^{2n}=(\cos \theta_j + \imi \sin \theta_j)^{2n}
\end{equation}
with $\theta_j=\frac{2j-1}{4n}\pi$. Comparing the real parts of both sides, we obtain
\begin{equation}
    0=\sum_{k=0}^{n} \binom{2n}{2k}
    (-1)^{n-k}(\cos \theta_j)^{2k} (\sin\theta_j)^{2(n-k)}, 
\end{equation}
which yields
\begin{equation}
    \sum_{k=0}^n \binom{2n}{2k}
    (\cot \theta_j)^{2k}(-1)^{n-k}=0.
\end{equation}
Here, we have used the fact that $\sin\theta_j \ne 0$ for $1\leq j \leq n$. We now consider an $n$th-order polynomial 
\begin{align}
    f(x)= \sum_{k=0}^n (-1)^{n-k} \binom{2n}{2k}
    x^k.
\end{align}
Then, we see that $(\cot \theta_j)^2$ $(j=1, \dots ,N)$ are 
distinct roots of the polynomial $f(x)$. Therefore, $f(x)$ is 
factorized as
\begin{align}
    f(x)= \prod_{j=1}^n \bigl[x-(\cot\theta_j)^2\bigr].
\end{align}
Comparing the coefficients of $x^k$, we obtain Eq.~\eqref{eq:sumformula}.
\end{proof}

Substituting Eq. \eqref{eq:sumformula} into Eq. \eqref{eq:sff3rd} yields the desired result Eq. \eqref{eq:Tr_H4^3}. Then substituting the obtained traces into the series expansion of $g(t,0)=|{\rm Tr} e^{{\rm i}t H_4}|^2/2^N$, we get Eq. \eqref{eq:g_t_short_time}. 

\subsection{Large $N$ behaviors \label{sec:sff-largeN}}
\begin{figure}
    \centering
    \includegraphics[width= \linewidth]{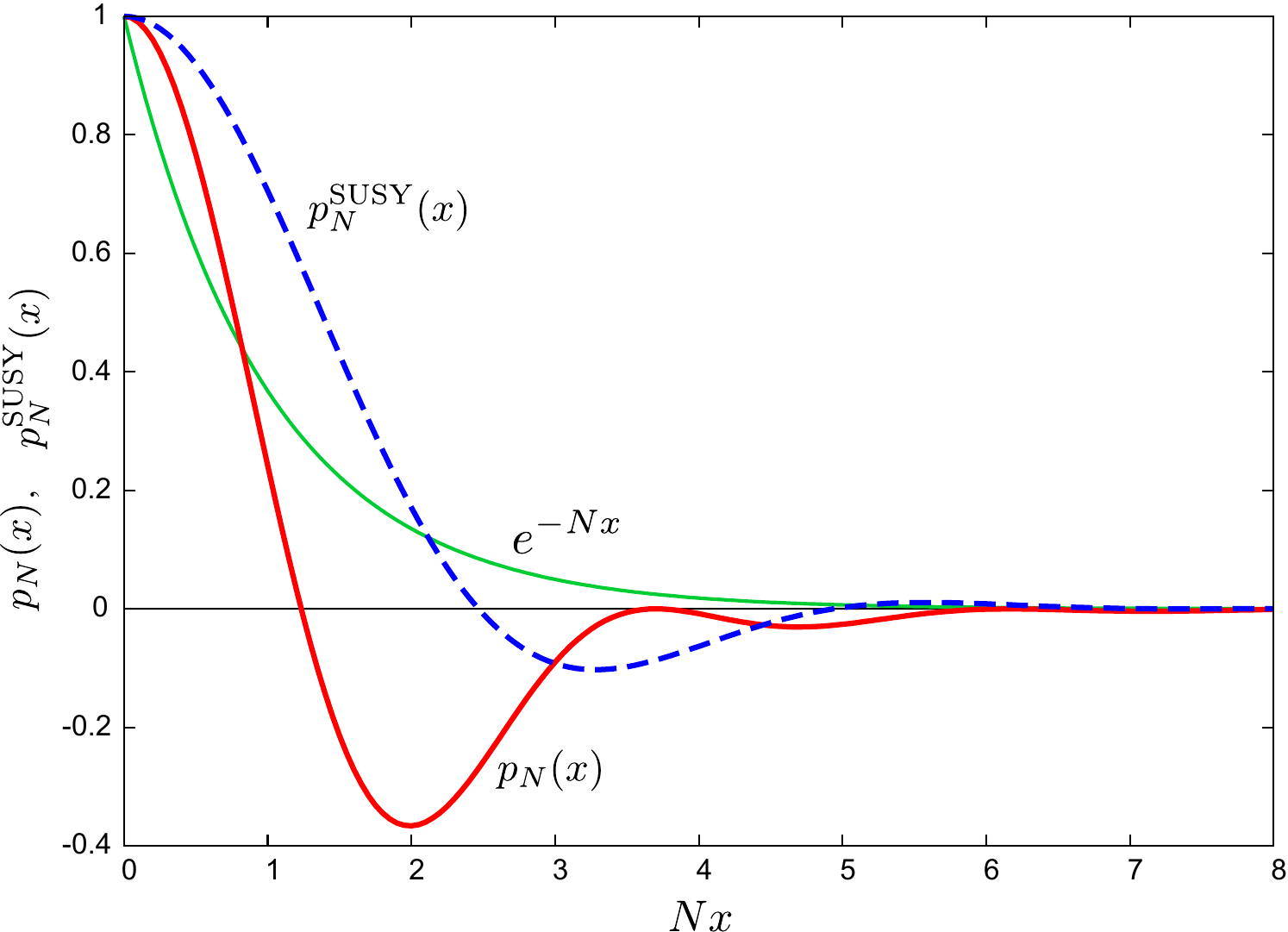}
    \caption{Function $p_N(x)$ (red solid line), $p^{\rm SUSY}_N(x)$ (blue dashed line), and their approximation $\tilde{p}_N(x)$ (green solid line) with $N=1000$.
    With this scaling of the axes, the shape of the graph does not change as long as $N$ is sufficiently large.
    This fact indicates that $p_{kN}(x)\sim p_{N}(kx)$ and $p^\text{SUSY}_{kN}(x)\sim p^\text{SUSY}_{N}(kx)$. }
    \label{fig:pfuncs}
\end{figure}

Next, we consider the large $N$ behaviors.
Using the Hubbard-Stratonovich (HS) transformation
\begin{equation}
    e^{\imi tA^2}=e^{\frac{\imi\pi}{4}}\sqrt{\frac{t}{\pi}} \int_{-\infty}^\infty e^{-\imi tx^2+2 \imi txA} dx,
\end{equation}
for Hermitian $A$ with the assumption that time $t$ is real, we obtain
\begin{widetext}
\begin{align}
    2^{-\frac{N}{2}}{\rm Tr}e^{\imi H_4 t} 
    =&e^{-\imi\frac{N(N-1)}{4}t} e^{\frac{\imi\pi}{4}} \sqrt{\frac{t}{\pi}}
    \frac{1}{2^\frac{N}{2}}\sum_{\sigma_1,\dots,\sigma_{N/2}=\pm 1}\int_{-\infty}^\infty 
    e^{-\imi tx^2} e^{\imi tx\cdot \frac{1}{\sqrt{2}} \sum_k \sigma_k \varepsilon_k} dx 
    \nonumber\\
    =&e^{-\imi\frac{N(N-1)}{4}t} e^{\frac{\imi\pi}{4}} \sqrt{\frac{2}{t\pi}}
    \int_{-\infty}^\infty 
    e^{-\imi\frac{2u^2}{t}} p_N(u) du ,
\end{align}
\end{widetext}
where 
\begin{equation}
    p_N(x)=\prod_{k=1}^{\frac{N}{2}}\cos(\epsilon_k x).
\end{equation}  
Here, we examine the property of $p_N(x)$.
We consider the single-particle DOS $D_1(\epsilon)$,
\begin{align}
    D_1(\epsilon)&=
    \frac{2}{N}\sum_{k=1}^{\frac{N}{2}} \delta(\epsilon -\epsilon_k)
    \nonumber \\
    &\simeq 2\int_0^{\frac{1}{2}} dk \delta (\epsilon-2\cot(k\pi)) \nonumber \\
    &=\frac{1}{\pi} \frac{1}{1+\epsilon^2/4}.
\end{align}
From this, we find that
\begin{align}
    |p_N(x)|&=e^{\sum_{k=1}^{k_{\rm max}} \log |\cos\epsilon_k x|} \nonumber \\
    &\simeq \exp\left(\frac{N}{2} \frac{1}{\pi} \int_0^\infty 
    \frac{4|x|}{\epsilon^2+4x^2} \log |\cos \epsilon| d\epsilon\right) \nonumber \\
    &= \left(\frac{1+e^{-4|x|}}{2}\right)^\frac{N}{2} \nonumber \\
    &\to e^{-N|x|} =: \tilde{p}_N(x) \quad (N\to \infty).\label{eq:p_exponentiallike}
\end{align}
The function $p_N(x)$ and its approximation $\tilde{p}_N(x)$ are shown in Fig.~\ref{fig:pfuncs}.
This result indicates that $p_N(x)$ decays almost exponentially over a range 
$[0,1/N]$.
Furthermore, 
our numerical results suggest an approximate scaling:
\begin{equation}
    p_{kN}(x) \sim p_N(kx),
\end{equation}
with positive integer $k$ for large $N$.
By the replacement of $N=kN_0$ with $N_0$ being the large positive integer, 
we can extract the scaling of $g(t,0)$ at large $t$ and 
$N$,
\begin{align}
    g(t,0)
    &=\frac{2}{\pi t}  \left|\int_{-\infty}^\infty 
    e^{-\imi\frac{2u^2}{t}} p_N(u) du \right|^2 \nonumber \\
    &\simeq \frac{8}{\pi t} \left(\int_{0}^\Lambda p_{N_0}(ku)du \right)^2 \nonumber \\
    &= \frac{8}{\pi t N^2} \left( N_0 
    \int_0^{k\Lambda} p_{N_0}(u)du\right)^2 \nonumber \\
    &=\frac{0.16}{t N^2}, \label{eq:sffnumerical}
\end{align}
where we have performed a numerical integration with $N_0=1000$ 
and sufficiently large cutoff $\Lambda$.
\soooo{The approximate functional form $p_N(x)\sim e^{-Nx}$ has been used solely to justify the truncation of the integral}
in the second line in Eq.~\eqref{eq:sffnumerical}
\soooo{and we use the correct form for the numerical integration in the third line}.
The result exhibits a power-law decay, which is in good agreement with the fully numerical result shown in Fig.~\ref{fig:sff-many} (a).

For $H_4^{\rm SUSY}$, similarly, we obtain
\begin{align}
    2^{-\frac{N}{2}} {\rm Tr} e^{\imi H^{\rm SUSY}_4 t}
    &=e^{\imi\frac{\pi}{4}}\frac{1}{\sqrt{\pi t}} \int^\infty_{-\infty} du\,
    e^{-\imi\frac{u^2}{t}}p_N^{\rm SUSY}(u),
    \label{eq:h4susy-sff-int}
\end{align}
where we have introduced
\begin{equation}
    p_N^{\rm SUSY}(x)=\prod_{k=1}^{\frac{N}{2}-1}\cos(\varepsilon_k x),
\end{equation}
which is also shown in Fig.~\ref{fig:pfuncs}.
This function is also approximated as $p_N^{\rm SUSY}(x) \simeq e^{-Nx}=\tilde{p}_N(x)$,
and we find 
an approximate scaling 
\begin{equation}
    p_{Nk}^{\rm SUSY}(x) \sim p_{N}^{\rm SUSY}(kx).
\end{equation}
Similarly to the analysis of $g(t,0)$, we assume this scaling relation and obtain
\begin{equation}
    g^{\rm SUSY}(t,0) \simeq \frac{1.9}{tN^2}
    \label{eq:gsusy_approx}
\end{equation}
by numerical integration. The fully numerical result for SFF of $H_4^\text{SUSY}$ and its approximation Eq.~\eqref{eq:gsusy_approx} are shown in Fig.~\ref{fig:sff-many} (b).

These results clearly show that the SFFs for $H_4$ and $H_4^{\rm SUSY}$ at large $t$ and 
$N$ are proportional to $1/N^2 t$.
\soo{The SFFs start to deviate from $0.16/N^2 t$ and $1.9/N^2 t$ at sufficiently late times.
The time at which the deviation occurs increases with $N$.}

\subsection{Spectral form factor at finite temperature \label{sec:sff-fintemp}}
Finally, we consider $g(t,\beta)$ and $g^\text{SUSY}(t,\beta)$ at 
finite temperature.
We introduce the general HS transformation,
\begin{equation}
    e^{-\alpha A^2} = \frac{e^{-\frac{\imi}{2}{\rm Arg}\alpha}}{\sqrt{|\alpha|\pi}}
    \int_{-\infty}^\infty dx\, e^{-\frac{x^2}{\alpha}+2iAx},
\end{equation}
for ${\rm Re}\,\alpha\geq 0$ with Hermitian $A$.
By the HS transformation, we obtain numerically tractable expressions for $Z_0(\beta)$ and $Z^\text{SUSY}(\beta)$ as
\begin{widetext}
\begin{equation}
    Z_0(\beta)=z(\beta) \sqrt{\frac{2}{\pi\beta}}2^{\frac{N}{2}} \quad \text{and} \quad 
    Z^\text{SUSY}(\beta)=z^\text{SUSY}(\beta) \frac{1}{\sqrt{{\pi\beta}}}2^\soo{\frac{N}{2}}
    \label{eq:partfuncs}
\end{equation}
with
\begin{equation}
    z(\beta)=\int_{-\infty}^\infty e^{-\frac{2x^2}{\beta}} p_N(x) dx \quad \text{and} \quad
    z\susy(\beta)=\int_{-\infty}^\infty e^{-\frac{x^2}{\beta}} p^{\rm SUSY}_N(x) dx.
    \label{eq:smallz_append}
\end{equation}
Using Eqs.~\eqref{eq:sff_append}, \eqref{eq:sff-susy}, \eqref{eq:partfuncs}, and \eqref{eq:smallz_append},
we obtain 
\begin{align}
    g_N(t,\beta)=\frac{1}{z\soo{(\beta)}^2} \frac{\beta}{\sqrt{t^2+\beta^2}}
    \bigg| \int_{-\infty}^\infty dx e^{-\frac{2x^2}{\beta-\imi t}}p(x) \bigg| ^2 
\end{align}
for $H_4$ and 
\begin{align}
    g^{\rm SUSY}_N(t,\beta)=\frac{1}{(z^{\rm SUSY}\soo{(\beta)})^2} \frac{\beta}{\sqrt{t^2+\beta^2}}
    \bigg| \int_{-\infty}^\infty dx e^{-\frac{x^2}{\beta-\imi t}}p^{\rm SUSY}(x) \bigg| ^2
\end{align}
\end{widetext}
for $H_4^{\rm SUSY}$.

\section{Thermodynamic properties\label{sec:app_thermodynamics}}
In this appendix, we 
detailed calculations of the thermodynamic properties of $H_4$ and $H_4^\text{SUSY}$.
For $H_4$, in the following, we use the partition function $Z_0(\beta)$ [Eq.~\eqref{eq:partfunc0}] to remove the constant shift of the total energy, $E_0$. 
In the high-temperature limit $\beta \to 0$,
the Gaussian distribution approaches $e^{-\frac{2u^2}{\beta}}\to \sqrt{\beta\pi/2}\delta(u)$.
Therefore, the free energy $F$ is given by 
\begin{equation}
    F(\beta)=-\frac 1 \beta \log Z_0(\beta) \sim -\frac{N T}{2} \log 2
    \quad (\beta\to 0),
\end{equation}
which is the expected behavior in the high-temperature limit.

Next, we consider the large $N$ limit. We evaluate $z(\beta)$ using the approximation $p_N(x)\sim \tilde{p}_N(x)$ and the asymptotic expansion of the error function as
\begin{align}
    z(\beta) \sim 2\int_0^\infty e^{-\frac{2x^2}{\beta}}e^{-Nx} dx
    =\frac{2}{N}+O\left( \frac{1}{N^3} \right). \label{eq:zapprox}
\end{align}
\soooo{Although the approximation we have used is not highly accurate as seen in Fig.~\ref{fig:pfuncs}, such a difference does not affect thermodynamic quantities.}
Combining Eq.~\eqref{eq:zapprox} with Eq.~\eqref{eq:partfuncs} we obtain
\begin{equation}
    F(\beta) \sim  -\frac {NT}{2}\log 2 \quad (N\to \infty),
\end{equation}
which indicates that the large $N$ limit leads to the same free energy as the high-temperature limit.
\begin{figure}
    \centering
    \includegraphics[width=\linewidth]{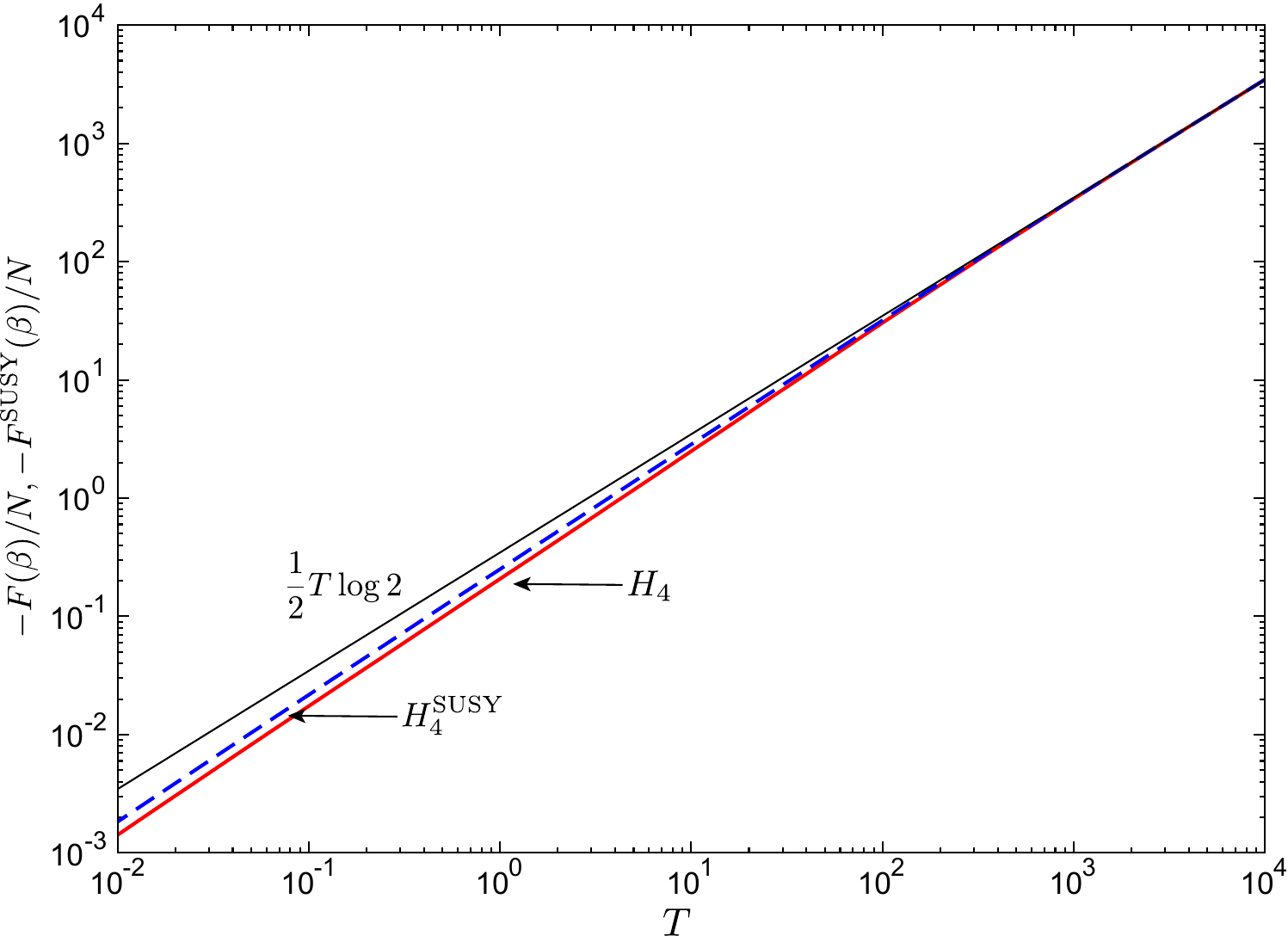}
    \caption{
    Free energy per particle for $H_4$ (red, solid line) and $H_4^\text{SUSY}$ (blue, dashed line) as functions of temperature with $N=36$. The value at large $N$ limit $\frac{1}{2}T\log 2$ is shown in black solid line. 
    }
    \label{fig:freeenergy}
\end{figure}

Similarly, we define the free energy for $H_4^\text{SUSY}$ as
\begin{align}
    F^\text{SUSY}(\beta)=-\frac{1}{\beta} \log Z^\text{SUSY}(\beta).
\end{align}
In the same way as $H_4$, the same expressions are obtained for the free energy for $H_4^{\rm SUSY}$.
Fully numerical evaluations of $F$ and $F^\text{SUSY}$
are shown in Fig.~\ref{fig:freeenergy},
which are in good agreement with the above analytical evaluations.
For $H_4$ and $H_4^\text{SUSY}$,
both in the high-temperature and large $N$ regimes, the entropy per degree of freedom is given by
\begin{equation}
    s=-\frac{1}{N}\left( \frac{\partial F}{\partial T} \right)
    _N\sim \frac{1}{2} \log 2 \quad (N\to \infty).
\end{equation}

\section{Dynamics \label{sec:app_dynamics}}
In this 
appendix, we derive the time evolution of $\gamma_j$ in the Heisenberg picture and 
present detailed calculations of the dynamical correlation functions.
We first note that the complex fermion operators $f_k^\pm$ and $g_k^\pm$ defined in Eqs. \eqref{eq:complexf} and \eqref{eq:complexg} satisfy 
\begin{align}
    [H_4, f_k^\pm] &=f_k^\pm (\pm\epsilon_k H_2 + \frac{1}{2} \epsilon_k^2), 
    \label{eq:commh4}\\
    [H_4^{\rm SUSY}, g_k^\pm] &= g_k^\pm (\pm 2\varepsilon_k H_{\rm free} +\varepsilon_k^2),
\end{align}
where $H_2$ and $H_\textrm{free}$ are defined in Eqs.~\eqref{eq:Ham_H2} and \eqref{eq:Ham_Hfree}.
These commutation relations imply
\begin{align}
    e^{\imi H_4 t}f_k^\pm e^{-\imi H_4 t} &= e^{-\frac{\imi}{2}\epsilon_k^2t} e^{\pm \imi \epsilon_k H_2 t} 
    f_k^\pm, \label{eq:gammat_append}\\
    e^{\imi H_4^{\rm SUSY} t}g_k^\pm e^{-\imi H_4^{\rm SUSY} t}
    & = e^{-\imi \varepsilon_k^2 t} e^{\pm 2 \imi\varepsilon_k H_{\rm free}t} g_k^\pm. \label{eq:timeevo_h4susy}
\end{align}
Since the Majorana 
and complex fermion operators 
are related by the Fourier transform
\begin{widetext}
\begin{align}
    f_k^\pm &=\frac{1}{2} (\psi_{2k-1} \mp \imi\psi_{2k})=\sqrt{\frac{1}{2N}} \sum_{j=1}^N 
    e^{\mp \imi(j-1)\theta_k} \gamma_j, \quad (1\leq k \leq \frac{N}{2}),\\
    g_k^\pm&=\frac{1}{2} (\chi_{2k-1} \mp \imi\chi_{2k})=\sqrt{\frac{1}{2N}} \sum_{j=1}^N 
    e^{\mp \imi(j-1)\vartheta_k} \gamma_j \quad (1\leq k \leq \frac{N}{2}-1),
\end{align}
$\gamma_j$ can be expressed in terms of the complex fermion operators 
as
\begin{align}
    \gamma_j=\sqrt{\frac{2}{N}} \sum_{s=\pm} \sum_{k=1}^{N/2} e^{ \imi s(j-1)\theta_k} f_k^\pm \label{eq:fourierf}
\end{align}
for $H_4$ and
\begin{equation}
    \gamma_j=\frac{1}{\sqrt{N}} \chi_0 + \frac{(-1)^{j-1}}{\sqrt{N}}\chi_{N/2} + 
    \sqrt{\frac{2}{N}} \sum_{s=\pm} \sum_{k=1}^{\frac{N}{2}-1} e^{ \imi s(j-1) \vartheta_k} g_k^\pm \label{eq:fourierg}
\end{equation}
for $H_4^{\rm SUSY}$.
Combining Eqs.~\eqref{eq:gammat_append}, \eqref{eq:timeevo_h4susy}, \eqref{eq:fourierf}, and \eqref{eq:fourierg}, we obtain the time evolution of $\gamma_j$ as
\begin{align}
    \gamma_j(t)&:= e^{\imi H_4 t} \gamma_j e^{-\imi H_4 t} = \sqrt{\frac{2}{N}} \sum_{s=\pm} \sum_{k=1}^\frac{N}{2}
    e^{\imi s (j-1)\theta_k} e^{-\frac{\imi}{2}\epsilon_k^2t} e^{ \imi s \epsilon_k H_2 t} 
    f_k^s, \label{eq:gamma_t_h4}\\
    \tilde{\gamma}_j(t)&:= e^{\imi H_4^{\rm SUSY} t} \gamma_j e^{-\imi H_4^{\rm SUSY} t} = \frac{1}{\sqrt{N}}\chi_0
    +\frac{(-1)^{j-1}}{\sqrt{N}}\chi_{N/2}  
    +\sqrt{\frac{2}{N}}
    \sum_{s=\pm} \sum_{k=1}^{\frac{N}{2}-1} 
    e^{\imi s(j-1)\vartheta_k}e^{-\imi \varepsilon_k^2 t}e^{2 \imi s \varepsilon_k H_{\rm free} t} 
    g_k^s.\label{eq:gamma_t_h4susy}
\end{align}

\subsection{Autocorrelation functions}\label{app:autoco}
We first consider the 
autocorrelation functions for $H_4$,
which is 
obtained by setting $l=m$ in Eq.~\eqref{eq:autocorr},
\begin{equation}
    G_{ll}(t,\beta)= {\rm Tr}[\rho \gamma_l(t) \gamma_l(0)],
    \label{eq:autocorr_append}
\end{equation}
where $\rho=\exp(-\beta H_4)/{\rm Tr}\exp(-\beta H_4)$. 
Substituting Eq.~\eqref{eq:gamma_t_h4} to Eq.~\eqref{eq:autocorr_append}, we obtain
\begin{align}
    G_{ll}(t) &=\frac{2}{NZ\soo{(\beta)}} {\rm Tr} \biggl[e^{-\frac{\beta}{2} (H_2)^2-\beta E_0} \sum_{k=1}^{N/2} 
    e^{-\frac{\imi}{2}\epsilon_k^2 t} e^{\imi\epsilon_k H_2 t\sigma_k}\biggr] \\
    &=\frac{2}{Nz\soo{(\beta)}} \sum_{k=1}^{N/2} \int_{-\infty}^\infty dx e^{-\frac{2}{\beta}x^2}
     e^{\imi x \epsilon_{k}} \prod_{k'\neq k} \cos\left(\frac{ \epsilon_k \epsilon_{k'} t}{2} + x\epsilon_{k'}\right) .
    \label{eq:autocorrh4}
\end{align}
with $\sigma_k=2 n_k -1$ and we have used the HS transformation.

For $H_4^{\rm SUSY}$, the 
two-point function is similarly defined as
\begin{align}
    G_{lm}^{\rm SUSY}(t)&:=  {\rm Tr}[\rho \tilde{\gamma}_l(t) \tilde{\gamma}_m(0)],
    \label{eq:autocorrh4susy}
\end{align}
where $\rho=\exp(-\beta H_4^{\rm SUSY})/{\rm Tr}\exp(-\beta H_4^{\rm SUSY})$. 
We now evaluate the autocorrelation function $G_{ll}^{\rm SUSY}(t)$. 
Substituting Eq.~\eqref{eq:gamma_t_h4susy} to Eq.~\eqref{eq:autocorrh4susy}, we obtain
\begin{align}
    G_{ll}^{\rm SUSY}(t)& =\frac{2}{N}+\frac{2}{Nz^{\rm SUSY}\soo{(\beta)}} \sum_{k=1}^{N/2-1} \int_{-\infty}^\infty dx e^{-\frac{x^2}{\beta}}
     e^{\imi x \varepsilon_{k}} \prod_{k'\neq k} \cos(\varepsilon_k \varepsilon_{k'} t + x \varepsilon_{k'}) ,
    \label{eq:autocorrh4susy_2}
\end{align}
where the first constant term originates from the Majorana zero modes contained in Eq.~\eqref{eq:gamma_t_h4susy}.

By numerical integration, we evaluate Eqs.~\eqref{eq:autocorrh4} and \eqref{eq:autocorrh4susy_2} as functions of time for several $\beta$. The results are presented in Fig.~\ref{fig:autocorr} (a) and (b), respectively.
We note that exponential decay of $|G_{ll}^{\rm SUSY}(t)-\frac{2}{N}|$ shown in Fig.~\ref{fig:autocorr}(b) implies the residual value $\frac{2}{N}$.

\subsection{Out-of-time-order Correlators}\label{app:otoc}
\subsubsection{Clean SYK model}
Next, we compute the OTOC for $H_4$. 
Substituting $\gamma_j(t)$ [Eq.~\eqref{eq:gamma_t_h4}] for Eq.~\eqref{eq:otoc} and using the commutation relation Eq.~\eqref{eq:commh4}, we 
perform the sum over $l$ and $m$, and obtain
\begin{align}
    F(t,\beta)=&-\frac{4}{N^2} \sum_{k_1\neq k_2} e^{-\frac{\beta}{4}(\epsilon_{k_1}^2+\epsilon_{k_2}^2)} {\rm Tr} [\rho e^{\frac{\beta}{2}(\sigma_{k_1}\epsilon_{k_1}+\sigma_{k_2}\epsilon_{k_2})H_2}
    e^{({\rm i}t-\frac{\beta}{4})\sigma_{k_1}\sigma_{k_2}\epsilon_{k_1}\epsilon_{k_2}}].
\end{align}
First, we consider the case with $\beta=0$. In this case, we have
\begin{align}
    F(t,0) =
    -\frac{4}{N^2} \sum_{k_1,k_2} \cos (\epsilon_{k_1} \epsilon_{k_2} t)
    +\frac{4}{N^2} \sum_{k_1} \cos (\epsilon_{k_1}^2 t).
    \label{eq:otoch4closed}
\end{align}
The first term can be computed analytically in the large-$N$ limit:
\begin{align}
    -\frac{4}{N^2} \sum_{k_1,k_2} \cos (\epsilon_{k_1} \epsilon_{k_2} t) 
    &
    \simeq-\frac{1}{\pi^2} \int_0^\infty d\epsilon_1 \int_0^\infty d\epsilon_2  \cos (\epsilon_1 \epsilon_2 t) 
    \frac{1}{1+(\frac{\epsilon_1}{2})^2} \nonumber
    \frac{1}{1+(\frac{\epsilon_2}{2})^2}  \\
    &=-\frac{2}{\pi}(\sin 4t \,{\rm Ci}\, 4t - \cos 4t \,{\rm si}\, 4t) \nonumber \\
    &=-\frac{2}{\pi}f(4t),
    \label{eq:otoc-leading-h4}
\end{align}
where $\text{si}\,x$ and $\text{Ci}\, x$ are sine and cosine integrals defined by
\begin{equation}
    \text{si}\,x= -\int_x^\infty \frac{\sin t}{t} dt
    \quad \text{and} \quad 
    \text{Ci}\,x = -\int_x^\infty \frac{\cos t}{t} dt,
\end{equation}
and 
\begin{equation}
    f(t):=\sin t \,{\rm Ci} \,t -\cos t \, {\rm si}\, t = \int_0^\infty \frac{\sin u}{u+t} du
\end{equation} is one of the auxiliary functions for the trigonometric integrals \cite{auxfunc}.
We now evaluate the second term in Eq. (\ref{eq:otoch4closed}). The estimate reads
\begin{align}
    \left| \frac{4}{N^2} \sum_{k_1=1}^{N/2} \cos (\epsilon_{k_1}^2 t) \right|  
    \simeq \frac{2}{N} \left| \int_0^\infty \frac{1}{\pi} \frac{\cos (\epsilon^2 t)}{1+(\frac{\epsilon}{2})^2}d\epsilon \right| 
    \leq \frac{2}{N} \int_0^\infty \frac{1}{\pi} \frac{1}{1+(\frac{\epsilon}{2})^2}d\epsilon
    =\frac{2}{N},
\end{align}
which means that this term is negligible in the large-$N$ limit. 
Thus we conclude that Eq.~\eqref{eq:otoc-leading-h4} dominates the behavior of $F(t,0)$. 
Furthermore, we obtain the asymptotic form
\begin{equation}
    F(t,0)=-\frac{1}{2\pi t}+ O\left(t^{-2}, \frac{1}{N}\right).
\end{equation}

Next, we consider the finite-temperature OTOC.
By the HS transformation,
\begin{align}
    e^{-\frac{\beta}{2}A^2}=\sqrt{\frac{\beta}{2\pi}}\int dx e^{-\frac{\beta}{2}x^2}e^{\imi x \beta A}
\end{align}
with $A=H_2-\frac{1}{2}(\sigma_{k_1}\epsilon_{k_1}+\sigma_{k_2}\epsilon_{k_2})$, we obtain the OTOC in the 
integral form as
\begin{align}
 F(t,\beta)=&-\frac{4}{z(\beta)N^2}\sum_{k_1\neq k_2} e^{-\frac{\beta}{8}
    (\epsilon_{k_1}^2+\epsilon_{k_2}^2)}
    \cos (\epsilon_{k_1} \epsilon_{k_2}t) \nonumber 
    \int_{-\infty}^\infty dx  e^{-\frac{2x^2}{\beta}}
    \prod_{k'\in K\backslash\{k_1,k_2\}}\cos (x \epsilon_{k'}),
\end{align}
where $K=\{1,2,\dots,\frac{N}{2}\}$.
This formulation enables us to evaluate $F(t,\beta)$ for large $N$,
and the result is shown in Fig.~\ref{fig:otoc_h4} (a) in the main text.

\soooo{Here, we provide details of the fitting of the OTOC. We perform the fitting 
using exponential and polynomial functions in the interval $[0.1\beta, 0.2\beta]$,
\begin{align}
 f_1(t)&=Ae^{\lambda' t/\beta}+B, \\
 f_2(t)&=a_0 +a_1 \left(\frac{t}{\beta}\right) + a_2 \left(\frac{t}{\beta}\right)^2,
\end{align}
where $A$, $\lambda'(=\lambda \beta)$, $B$, $a_0$, $a_1$, and $a_2$ are the fitting parameters.
The results for $H_4$ are shown in Table~\ref{table:fitting} and Fig.~\ref{fig:fitting}.
We find that the exponential fit is better because a simple proportional relation between the fitted exponent and temperature is found, which is predicted for quantum many-body chaotic systems. On the other hand, for the polynomial fit, we find no simple relation between the fitted parameters 
and temperature.
}
\begin{table}
    \caption{Fitting parameters for OTOC for $H_4$. \label{table:fitting}}
    \begin{ruledtabular}
    \begin{tabular}{c|ccc|ccc}
     $\beta$ &  $A$ & $\lambda'(=\beta\lambda)$ & $B$ & $a_0$ & $a_1$ & $a_2$  \\ \hline
    2.0 & 0.02051  & 3.791 & -0.2058 & -0.1844 & 0.05887 & 0.2631\\
    1.0 & 0.01751  & 4.342 & -0.2933 & -0.2745 & 0.05444 & 0.3203\\
    0.5 & 0.01339 & 4.801 & -0.3936 & -0.3788 & 0.03659 & 0.3208\\
    0.2 & 0.008131 & 5.247 & -0.5311 & -0.5218 & 0.01969 & 0.2491 \\
    0.1 & 0.005102 & 5.483 & -0.6281 & -0.6221 & 0.01111 & 0.1768 \\
    0.05 & 0.003011 & 5.652 & -0.7124 & -0.7088 & 0.005931 & 0.1138
    \end{tabular}
    \end{ruledtabular}
\end{table}

\begin{table}
    \caption{Fitting parameters for OTOC for $H_4^{\rm SUSY}$. \label{table:fittingsusy}}
    \begin{ruledtabular}
    \begin{tabular}{c|ccc}
     $\beta$ &  $A$ & $\lambda'(=\beta\lambda)$ & $B$ \\ \hline
    0.5 & 0.01774 & 4.354 & -0.2953 \\
    0.2 & 0.01199 & 4.937 & -0.4298  \\
    0.1 & 0.008119 & 5.258 & -0.5342  \\
    0.05 & 0.005122 & 5.479 & -0.6315
    \end{tabular}
    \end{ruledtabular}
\end{table}

\subsubsection{Clean SUSY SYK model}
Similarly to $H_4$, we can calculate the OTOC for the clean SUSY SYK model.
We obtain the finite-temperature OTOC for $H_4\susy$ as 
\begin{align}
    F^{\rm SUSY}(t,\beta) =& \frac{1}{N^2} \sum_{l,m=1}^{N/2}{\rm Tr} [\rho^\frac{1}{4} \tilde{\gamma}_l(t) \rho^\frac{1}{4} \tilde{\gamma}_m(0) \rho^\frac{1}{4}  \tilde{\gamma}_l(t) 
    \rho^\frac{1}{4} \tilde{\gamma}_m(0)] \nonumber \\
    =&-\frac{4}{N^2} \sum_{k_1\neq k_2} 
    e^{-\frac{\beta}{2}(\varepsilon_{k_1}^2+\varepsilon_{k_2}^2)}
    {\rm Tr}
    \bigl[ \rho
    e^{\beta(\sigma_{k_1}\varepsilon_{k_1}+\sigma_{k_2}\varepsilon_{k_2})H_{\rm free}}
    e^{(2\imi t-\frac{\beta}{2})\sigma_{k_1}\sigma_{k_2}\varepsilon_{k_1}\varepsilon_{k_2}}
    \bigr] \nonumber \\
    &-\frac{8}{N^2}\sum_{k_1=1}^{\frac{N}{2}-1} {\rm Tr}
    [\rho e^{-\frac{\beta}{2}(\varepsilon_{k_1}^2-2\sigma_{k_1}\varepsilon_{k_1}H_{\rm free})}].
\end{align}

As done for the OTOC of $H_4$ at $\beta=0$, we obtain 
\begin{align}
    F^{\text{SUSY}}(t,0) &= -\frac{4}{N^2} \sum_{k_1,k_2} [\cos (2\varepsilon_{k_1} \varepsilon_{k_2} t) -
    \delta_{k_1,k_2} \cos (2\varepsilon_{k_1}^2 t)] -\frac{4(N-2)}{N^2}\nonumber \\
    &\simeq -\frac{2}{\pi}(\sin 8t \,{\rm Ci}\, 8t - \cos 8t \,{\rm si}\, 8t) +O\bigg(\frac{1}{N}\bigg)\nonumber \\
    &=-\frac{1}{4\pi t}+O\left(t^{-2}, \frac{1}{N}\right).
\end{align}

Using the HS transformation, we obtain the OTOC for $H_4^{\rm SUSY}$
at finite temperature in the integral form as

\begin{align}
 F^\text{SUSY}(t,\beta)=&-\frac{4}{z\susy (\beta)N^2}\sum_{k_1\neq k_2} e^{-\frac{\beta}{4}
    (\varepsilon_{k_1}^2+\varepsilon_{k_2}^2)}
    \cos (2\varepsilon_{k_1} \varepsilon_{k_2}t)
    \int_{-\infty}^\infty dx  e^{-\frac{x^2}{\beta}}
    \prod_{k'\in K\susy \backslash\{k_1,k_2\}}\cos (x \varepsilon_{k'})
    \nonumber \\
    &-\frac{8}{z\susy (\beta) N^2}\sum_{k_1}e^{-\frac{\beta}{4}\varepsilon_{k_1}^2}
    \int_{-\infty}^\infty dx e^{-\frac{x^2}{\beta}} \prod_{k' \in K\susy \backslash \{k_1\}} \cos (x \varepsilon_{k'})
\end{align}
with $K^\text{SUSY}=\{1,\dots,\frac{N}{2}-1\}$.
The numerical evaluation of $F^\text{SUSY}(t,\beta)$ 
is shown in Fig.~\ref{fig:otoc_h4} (b) and the results of exponential fitting are summarized in Table~\ref{table:fittingsusy}, which are 
quite similar to those 
of $F(t,\beta)$ for $H_4$.
\end{widetext}

\begin{figure}
    \centering
    \includegraphics[width=\linewidth]{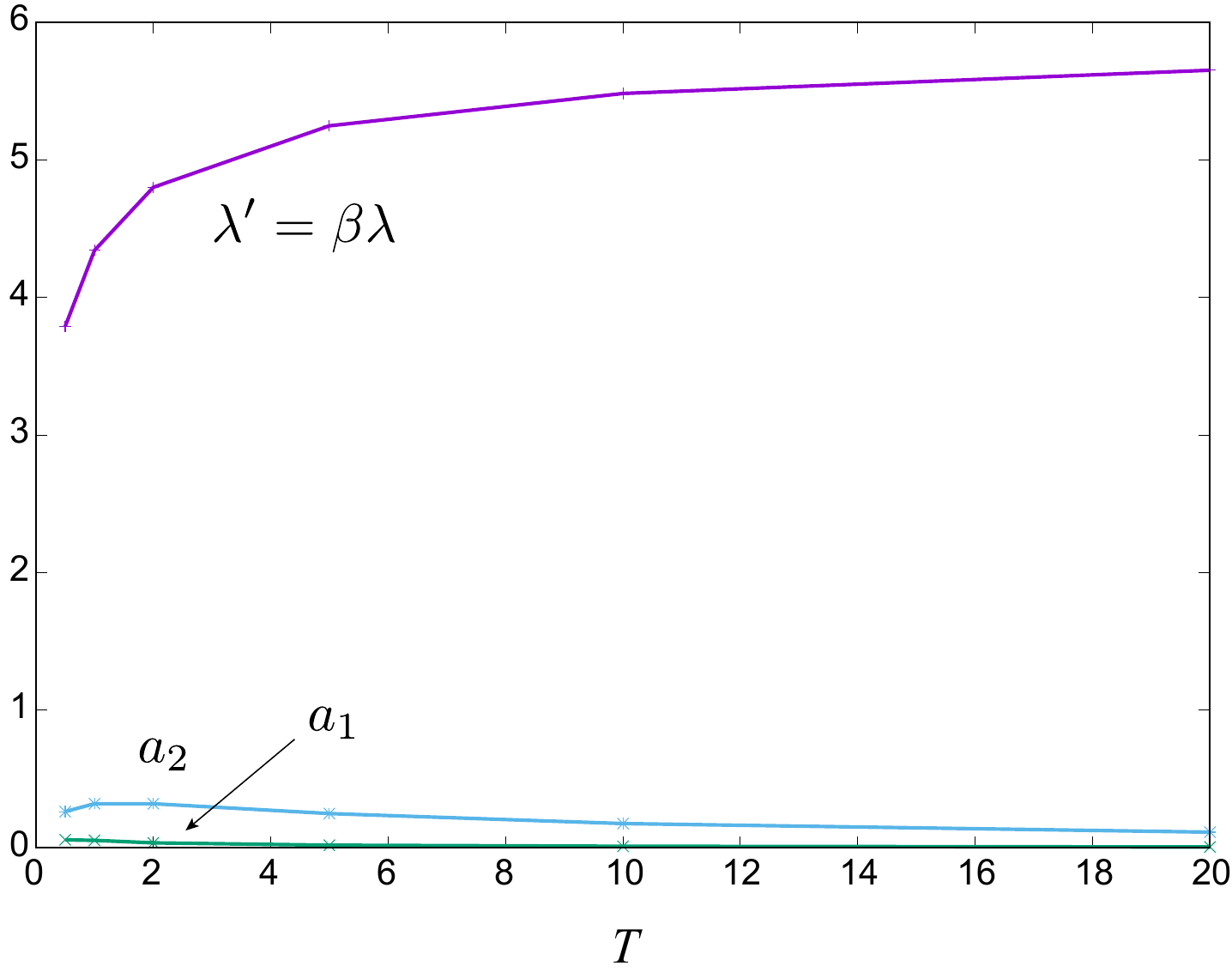}
    \caption{Fitting parameters as functions of the temperature $T$.}
    \label{fig:fitting}
\end{figure}

\subsubsection{Quadratic clean SYK model $H_2$}
To see the fact that the scrambling behavior of the OTOC is characteristic in four-body SYK-like systems, we calculate the OTOC for $H_2$ at 
the infinite temperature.
The OTOC for $H_2$ is obtained as
\begin{align}
    F_{\rm quad}(t)&=\frac{1}{N^2} \sum_{l,m} 2^{-\frac{N}{2}}
    \text{Tr} [\bar{\gamma}_l(t)\bar{\gamma}_m(0)\bar{\gamma}_l(t)\bar{\gamma}_m(0)] \nonumber \\
    &=-1+\frac{2}{N}
\end{align}
with $\bar{\gamma}_l(t)=e^{\imi H_2 t}\gamma_l e^{-\imi H_2 t}$, which does not depend on time.

\subsubsection{Free fermion chain}
As another example of a noninteracting system, we consider the OTOC for a free-fermion chain. 
The Hamiltonian is given by
\begin{equation}
    H_{\rm 1d} = 
    -\sum_{i=1}^N c_i^\dagger c_{i+1} + \text{H.c.},
\end{equation}
where $c_i$ ($c_i^\dagger$) is an annihilation (creation) operator for fermion at site $i$,
we have set the hopping integral to be unity, and periodic boundary conditions are imposed: $c_{N+1}=c_1$ and $c_{N+1}^\dagger=c_1^\dagger$.
We define the OTOC for this complex fermion system 
in a manner similar to that described in Ref.~\cite{iyoda2018}, as 
\begin{align}
    F_{\rm 1d}(t)=\frac{1}{N^2} \sum_{l,m=1}^N 2^{-N}
    \text{Tr}[c_l^\dagger(t)c_m^\dagger(0) c_l(t) c_m(0)],
    \label{eq:otoccomplex}
\end{align}
where $c_l^{(\dagger)}(t) = e^{\imi H_{\rm 1d}t}c_l^{(\dagger)} e^{-\imi H_{\rm 1d} t}$.
Equation~\eqref{eq:otoccomplex} is calculated as
\begin{align}
    F_{\rm 1d}(t)=-\frac{1}{4}\left(1-\frac{1}{N}\right),
\end{align}
which does not exhibit scrambling behavior either.

\bibliography{refs}
\end{document}